\documentclass{aa}
\usepackage{graphicx,txfonts,natbib}

\newcommand{\teff}{\mbox{${T}_{\rm eff}$}}
\newcommand{\logg}{\mbox{${\log g}$}}
\newcommand{\tgdiagram}{(\teff,\logg) diagram}

\begin{document}

\title{CoRoT's view of newly discovered B-star pulsators: results for 358
candidate B pulsators from the initial run's exoplanet field data\thanks{The
CoRoT space mission was developed and is operated by the French space agency
CNES, with the participation of ESA's RSSD and Science Programmes, Austria, Belgium,
Brazil, Germany, and Spain.  All frequency tables, including the identification
of combination frequencies, are only available as online material}}

\subtitle{}
\author{P.~Degroote\inst{1}           
\and
C.~Aerts\inst{1,2}
\and
M.~Ollivier\inst{3}
\and
A.~Miglio\inst{4}\thanks{Postdoctoral Researcher, Fonds de la Recherche
Scientifique - FNRS, Belgium} 
\and
J.~Debosscher\inst{1}
\and
J.~Cuypers\inst{5}
\and
M.~Briquet\inst{1}\thanks{Postdoctoral Fellow of the Fund for Scientific
Research, Flanders}
\and
J.~Montalb\'an\inst{4} 
\and
A.~Thoul\inst{4} 
\and
A.~Noels\inst{4} 
\and
P.~De~Cat\inst{5}
\and
L.\,Balaguer-N\'u\~nez\inst{6}
\and
C.~Maceroni\inst{7}
\and
I.~Ribas\inst{8}
\and
M.~Auvergne\inst{9}
\and
A.~Baglin\inst{9}
\and
M.~Deleuil\inst{10}
\and
W.~Weiss\inst{11}
\and
L.~Jorda\inst{10}
\and
F.~Baudin\inst{3}
\and
R.~Samadi\inst{9}
}

\offprints{P.~Degroote}

\institute{Institute of Astronomy - K.U.Leuven, Celestijnenlaan 200D, B3001 Leuven, Belgium
\and Department of Astrophysics, University of Nijmegen, PO Box 9010, 6500 GL Nijmegen, The Netherlands
\and Institut d'Astrophysique Spatiale (IAS), B\^atiment 121, F-91405, Orsay Cedex, France
\and Institut d'Astrophysique et de G\'eophysique Universit\'e de Li\`{e}ge, All\'{e}e du 6 Ao\^{u}t 17, B 4000 Li\`{e}ge, Belgium
\and Koninklijke Sterrenwacht van Belgi\"e, Ringlaan 3, 1180 Brussels, Belgium
\and Departament d'Astronomia i Meteorologia, Universitat de Barcelona, Av. Diagonal, 647, 08028 Barcelona, Spain
\and INAF-Osservatorio di Roma, via Frascati-33, Monteporzio Catone (RM), Italy
\and Institut de Ci\`encies de l'Espai (CSIC-IEEC), Campus UAB, Facultat de Ci\`encies, Torre C5 parell, 2a pl, 08193 Bellaterra, Spain
\and LESIA, UMR8109, Universit\'e Pierre et Marie Curie, Universit\'e Denis Diderot, Observatoire de Paris, 92195 Meudon Cedex, France
\and LAM, UMR 6110, CNRS/Univ. de Provence, 38 rue F. Joliot-Curie, 13388 Marseille, France
\and Department of Astronomy, University of Vienna, T\"urkenschanzstrasse 17, A-1180 Wien, Austria
}

   \date{Received 19 February 2009; accepted 26 May 2009}
   \authorrunning{P. Degroote et al.}
   \titlerunning{Asteroseismology of B stars in CoRoT's Initial Run Exofield}

  \abstract
  {We search for new variable B-type pulsators in the CoRoT data assembled primarily for
  planet detection, as part of CoRoT's Additional Programme.}
  {We aim to explore the properties of newly discovered B-type pulsators from the
  uninterrupted CoRoT space-based photometry and to compare them with known members of the $\beta\,$Cep and slowly pulsating B star (SPB) classes.}
  {We developed automated data analysis tools that include algorithms for
  jump correction, light-curve detrending, frequency detection, frequency 
  combination search, and for frequency and period spacing  searches.} 
  {Besides numerous new, classical, slowly pulsating B stars,
  we find evidence for a new class of low-amplitude B-type pulsators between the SPB and $\delta$\,Sct instability strips, with a very broad range of frequencies and low amplitudes, 
as well 
  as several slowly pulsating B stars with residual excess power at frequencies typically
  a factor three above their expected g-mode frequencies. }
  {The frequency data we obtained for numerous new B-type pulsators
  represent an appropriate starting point for further theoretical analyses of these stars, once
  their effective temperature, gravity, rotation
  velocity, and abundances will be derived spectroscopically in the framework of an ongoing FLAMES survey at the VLT.}
  \maketitle
%

\section{Introduction}
In the forthcoming months and years, the French-European Space mission CoRoT \citep[Convection, Rotation and planetary Transits][]{auvergne2009}, 
will measure brightness variations of tens of thousands
of different stars in its exoplanet programme, with a precision typically two
orders of magnitude better than we can achieve from ground-based observations
today. This means that we are able to extend the sample of known pulsating
stars to class members with fainter apparent magnitude. The study of stellar
oscillations from the exoplanet data of the mission is part of the Additional
Programme \citep{weiss2004}.

In this paper, we focus on CoRoT's Initial Run in the anticentre direction of
the Galaxy (IRa01). We first describe our methods of analysis, as future runs will also be addressed with
the same methodology.
During IRa01, which lasted some 55 days, the CoRoT satellite was pointed at
a field in the galactic plane, towards the galactic anticentre
($\alpha\approx06$\,h\,$44$\,m, $\delta\approx-01\degr\,12\arcmin$). We are
focusing on the $\beta$\,Cep stars \cite[e.g.][]{stankov2005} and slowly pulsating
B stars \citep[SPBs, ][]{waelkens1991} discovered in this field. A short
descriptive overview of the main properties of these two classes is listed in
Table\,\ref{tbl:overview}. There is a small overlap between their instability regions, where hybrid pulsators of spectral type near B3, exhibiting both high-order g and low-order p
modes, are predicted but not yet observed. On the other hand, these two types of modes have been observed in hotter $\beta$\,Cep stars ($\nu$~Eri: \citealt{handler2004}; 12 Lac: \citealt{handler2006}; $\gamma$ Peg: \citealt{chapellier2006}). The mechanism that drives the pulsations is the $\kappa$ mechanism
\citep{dziembowski1993} because of the Fe-group opacity bump at $T\sim 2\times 10^5$\,K, which means that the excitation of modes in those stars is sensitive to the behaviour of opacity in the stellar interior and to the abundance of Fe-group elements.

\begin{table}
\centering
\caption{Overview of the main properties of $\beta$\,Cep and SPB
pulsators.}\label{tbl:overview}
\begin{tabular}{rcccc}\hline\hline
Star type & \teff (K)     & Mass ($M_\odot$) & $f$ (d$^{-1}$) & main modes\\
\hline
SPB            & 10\,000 - 22\,000 & 3 - 8  & 0.3 - 1     & g modes\\
$\beta$\,Cep & 20\,000 - 30\,000 & 7 - 20 & 4 - 12      & p modes\\\hline
\end{tabular}
\end{table}

It would be a huge loss of resources to start analysing all stars in the CoRoT
database only to study the B-type pulsators. Therefore, we resort to the results
obtained in the framework of the variability 
classification programme described in \citet{debosscher2009}, hereafter termed CVC,
which extracts a basic number of relevant parameters from each
light curve to compare them with training sets of stars of known type.

In this paper, we focus on all light curves given a label of either $\beta$
Cep or an SPB star. 
We also include the $\delta$\,Sct and $\gamma$~Doradus stars to
limit the red edge of the SPB instability strip and to discuss possible overlap
between $\beta$\,Cep and $\delta$\,Sct stars, on the one hand, 
and between SPB and $\gamma$~Doradus stars on the other, in
terms of light curve properties (e.g., 
frequency values). In total, 540 light curves are extracted. We emphasise
that the results from the CVC classifier are only used to select the initial sample of 
stars from the CoRoT database.

\section{Description of the dataset and pre-processing}\label{descr}

\begin{figure}
\includegraphics[width=\columnwidth]{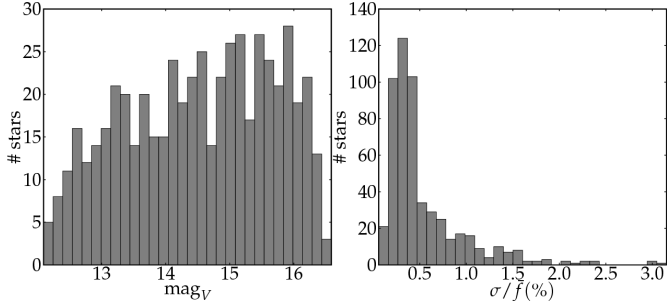}
\caption{Distribution of visual magnitudes (\emph{left}) and the ratio of the
light curve's standard deviation to the mean of the flux
(\emph{right}).}\label{fig:histograms}
\end{figure}

All datasets are taken over a period of
$55$\,d. Two time samplings occur, one measurement every $\sim$\,$32$\,s or one
every $\sim$\,$512$\,s. This sampling is not always constant for one light curve,
it may change from low temporal resolution to high temporal resolution during
the observing run. The magnitudes of the stars in the sample lie between 12.1
and 16.6 and vary over different scales (Fig.\,\ref{fig:histograms}).

\begin{figure}
\includegraphics[width=\columnwidth]{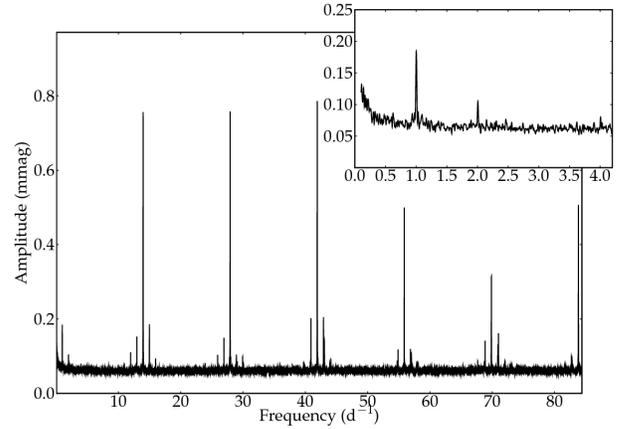}
\caption{Averaged Fourier periodogram of 100 non-variable stars. The inset is a
zoom on the low-frequency region. A low-frequency powerlaw trend is present, as are
the 1 and 2 day alias and the orbital frequency with many of its harmonics and
day aliases.}\label{fig:instr_eff}
\end{figure}

To have an idea of the different kinds of instrumental effects we can expect, as
well as their impact on the analysis, we analysed 100 constant stars
and averaged their Fourier periodogram. They were selected with the criterion
that their dominant frequency has a $p$-value
$>$\,$0.01$. From Fig.\,\ref{fig:instr_eff} we can identify at least three
different groups of features: a low-frequency powerlaw trend due to
instrumental drifts (red noise), which becomes negligible around
$0.4$\,d$^{-1}$, peaks at 1, 2, and 4\,d$^{-1}$ and peaks connected with the orbital
frequency due to temperature changes. We refer to \citet{auvergne2009} for a discussion of the origin
of these instrumental peaks and to \citet{samadi2007} for a description of the CoRoT data pipeline
and treatment of data during the passage of the South Atlantic Anomaly.

The most variable feature is the low-frequency powerlaw trend, because of which we introduced
a pre-processing step to minimise its potential influence. Two phenomena add up
to this trend: discontinuities in the light curve and long-term effects, possibly
of instrumental origin. It is impossible to decide whether or not these longterm
trends are real variations in the star's light curve, but because we do encounter
trends suspiciously linear or exponential, we choose to eliminate them. This
also means that we effectively remove possible oscillating behaviour and
intrinsic trends on the timescale of the time series itself.

\begin{figure}
\includegraphics[width=\columnwidth]{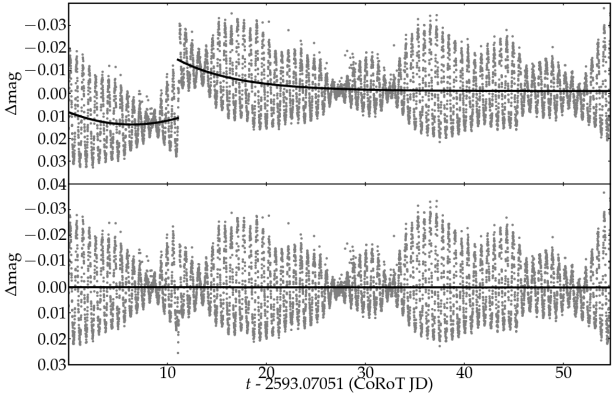}
\caption{Correction of one discontinuity in the light curve of SPB candidate
102815542. The first part and second parts were detrended using a quadratic
polynomial and exponential decay model, respectively.}\label{fig:jump}
\end{figure}

Long-term trends are removed in the following iterative way. First, jumps are
detected by sliding two adjacent bins over the lightcurve. Physically, we
only expect to find a gradual, continuous rise or decay in the measured flux. To accomodate
for the dependence of amplitude and present pulsations, we compare the difference between
the average of the bins to the overall difference. If at any
point the difference is too high to be considered continuous, as defined by the entire light curve,
we identify the point as a discontinuity,
perform a piecewise detrending of the two separate parts of the lightcurve and
try to detect other jumps. If no more jumps are detected, the piecewise
detrending is redone using the original light curve and the discontinuities
previously detected. There is some arbitrariness in the choice of parameters
(bin width, threshold value etc.), so they were fixed empirically. For the
piecewise detrending, the best fit of three simple models was chosen for each
part; either a linear, a quadratic, or an exponential trend. An example is
provided in Fig.\,\ref{fig:jump}.

\section{Frequency analysis methodology}

\subsection{Basic treatment}

To analyse all jump-corrected and detrended light curves, we describe the flux $F(t)$ as
\begin{equation}F(t_i) = \mu + C + \sum_{i=1}^n 
A_i\sin(2\pi(f_it+\phi_i)) + \epsilon_i,
\label{eq:observed}\end{equation}
where $\epsilon_i$ is a Gaussian noise component, $\mu$ the
observed average flux level, and $C$ a usually small correction parameter on
$\mu$. In this model, there are $n$ nonlinear parameters $f_i$, for which the
values are estimated using the Lomb-Scargle version of the discrete Fourier transform \citep{lomb1976,scargle1982}. The linear parameters $A_i$ and $\phi_i$
describe the amplitude and the phase of the corresponding pulsation,
respectively, and are estimated by ordinary least squares regression.
The error on the frequency determination (and other parameters) is calculated
according to \citet{montgomery1999}. These error determinations underestimate
the real error because of correlation effects in the data. In our sample, this factor lies between 1 and 4 for
the first frequency, and decreases for subsequent frequencies
\citep{schwarzenberg2003}.

Traditionally, the frequency acceptance criterion
of \citet{breger1993}, using a signal-to-noise ratio (SNR) lower limit of
$\mbox{SNR}=4$ in the amplitude periodogram for a peak to be considered genuine,
has been proven to be very successful, since very few frequencies
identified this way (if at all) had to be refuted afterwards. However, there is no rule
about how large one has to choose the interval in the periodogram to use for
noise calculation. Second, this method is not suited to detecting regions of
power excess, either due to densely packed frequencies or stochastically excited
frequencies. Third, related to the previous argument, the SNR method does not
guarantee that, if a peak is considered as due to noise, there is no other,
possibly even smaller, peak in the periodogram that is \emph{not}
noise. Therefore, the SNR method is not fully objective in its use.

Instead of using the classical SNR criterion, we make an assumption on what the
datasets should look like when devoid of instrumental effects and variability
from the star itself. A reasonable (yet not perfect) assumption is that we are
dealing with white Gaussian noise and that the number of datapoints is large.
\citet{degroote2009} provide an extensive discussion of the deviation from white
Gaussian noise and have shown that it can be ignored in the frequency analysis,
as long as a correction factor for correlated data is used.
Under this assumption, the distribution of the Scargle power spectrum (normalised with the total variance of the data) belongs to the exponential family \citep{scargle1982},
\begin{equation}P(z) = \exp(-z)\label{eq:exp},\end{equation}
and thus the noise level is located at a height of $z=1$. In reality, we never
test a single frequency, but cover a wide range of possible frequencies, between
0 and the highest detectable frequency, the Nyquist frequency $f_{Ny}$. In theory, this
number is well-defined regardless of the time spacing \citep{eyer1999}. In practice, however, this frequency is usually too high to be of any use \citep[e.g.][]{koen2006}.
A better practical guess for the Nyquist frequency in uneven time sampling is obtained
via
\begin{equation}f_{Ny}=\frac{1}{2 \Delta t},\label{eq:nyquist}\end{equation}
with $\Delta t$ in our case the value of the larger timesteps.

\begin{figure}
\includegraphics[width=\columnwidth]{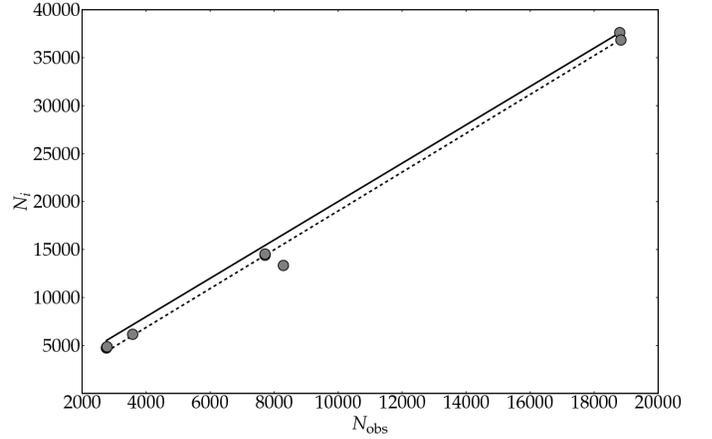}
\caption{Result of a simulation study to determine the number of independent
frequencies in the CoRoT datasets (\emph{sold line}: evenly spaced case,
$N_i=2N_{obs}$, \emph{dotted line}: fit through the simulation
results).}\label{fig:indep_freqs}
\end{figure}

To account for the simultaneous testing of all frequencies up to $f_{Ny}$, we
apply the Bonferroni correction \citep{scargle1982} to Eq.\,(\ref{eq:exp}):
\begin{equation}P_B(z) = 1 - [1 - \exp(z)]^{N_i}.\label{eq:faf}\end{equation}
For equally spaced data, theory predicts that the number of independent
frequencies $N_i$ between 0 en $f_{Ny}$ scales with the number of observations
$N_{obs}$ as
\begin{equation}
N_i=2N_{obs}. \label{eq:ni}
\end{equation}
In the case of unevenly spaced data, this parameters $N_i$ loses its meaning
\citep[e.g.][]{frescura2008}. However, a simulation study shows that expression (\ref{eq:ni}), combined with definition
(\ref{eq:nyquist}), turns out to be a useful and conservative approximation of
the true number of independent frequencies. For
eight different time templates taken from a CoRoT light curve, we generated
10\,000 light curves consisting of white Gaussian noise. Next, we calculated the
height $z$ of the maximum peak in the Scargle periodogram. These were combined,
and a false alarm function of the form (\ref{eq:faf}) fitted through the
data, to determine the free parameter $N_i$. The result can be seen in
Fig.\,\ref{fig:indep_freqs}, and shows that the CoRoT data allow the use of Eq. (\ref{eq:ni}).

Although the sampling rate is usually much higher than strictly needed, we choose not to rebin the data or to convert the flux to
magnitude, because such manipulations alter the shape of a light curve.

\subsection{Search for combination frequencies}

In particular cases, model (\ref{eq:observed}) can be expanded further to
lower the number of dependent parameters and thus simplify the model in
terms of $f_i$. Neglecting this effect implies imposing a linear model on all observed light curves, thus treating every found pulsation frequency as an independent mode. Identifying dependent frequencies can be seen as a basic correction of the linear model for nonlinear effects:
\begin{equation}
\begin{array}{ll}
\displaystyle F(t) = \mu + C 
&\displaystyle + \sum_{k=1}^{n_f}A_k\sin[2\pi(f_kt_i+\phi_k)]\\
\displaystyle &\displaystyle + \sum_{l=1}^{m_f}A_l\sin[2\pi(f_lt_i + \phi_l)]\\
\end{array}\label{eq:observed_expanded}
\end{equation}
with
\[f_l = n_l^1f_l^1 + n_l^2f_l^2 + n_l^3f_l^3,\]
where $n_f<n$ and $f_l$ represent combination
frequencies. This might come from nonlinear effects, e.g., nonlinear coupling between modes
\citep[e.g.]{buchler1997,handler2006,degroote2009}.
In model (\ref{eq:observed_expanded}), the linear parameters
still have to be fitted, but are sometimes expected to be in phase or in
anti-phase with each other. In contrast to the window frequencies, combination
frequencies and harmonics are not 
artificially introduced, because they have a real
physical interpretation. It is thus important to derive them.

The search for combination frequencies is done in the
following way: for each frequency, the harmonics are identified using all
empirically found frequencies with higher amplitude than the one under
investigation. Those are then excluded from the list of independent
frequencies. For the remaining frequencies, all second and third order
combinations are tested, again using only the frequencies with higher amplitude,
and not part of a combination so far (in order to avoid higher order
`combinations of combinations'). To keep the parameter space manageable, frequencies of order $n$ are only
considered if at least one of order $(n-1)$ is previously identified. We are limited by the total time of observations, so
there is some ambiguity in the choice of possible deviation from the exact
combination. Because the Rayleigh limit of $L_R=1/T$ is a natural measure of
maximum uncertainty on the frequency determination, we consider this a
reasonable value to test the presence of combinations.

\subsection{Search for period and frequency spacings}

\begin{figure}
\centering\includegraphics[width=.9\columnwidth]{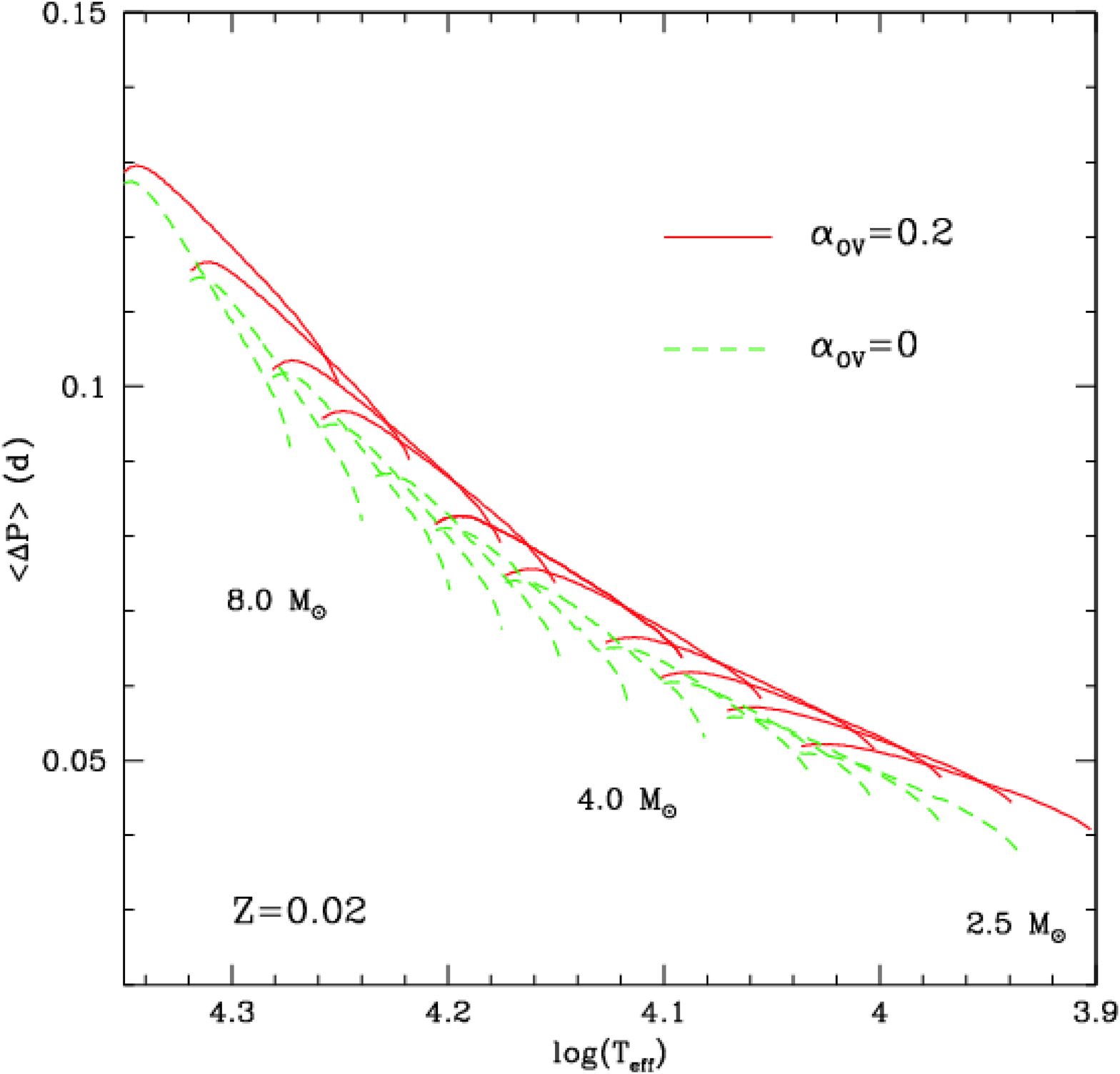}
 \caption{Theoretically predicted asymptotic $\ell=1$ period spacing of high-order g modes ($\Delta P$) as a function of the effective temperature. $\Delta P$ is computed for main-sequence models with masses between 2.5 and 8 $M_\odot$, with (full line) and without (dashed line) overshooting from the convective core. The heavy-element mass fraction assumed in the models is $Z=0.02$.}\label{fig:deltap}
 \end{figure}

The variability detected in SPB stars is interpreted as stemming from low degree, high-order modes whose periods are known to be sensitive to the structure of the stellar core.
The first-order asymptotic approximation developed by \citet{tassoul1980} shows that the period spacing ($\Delta P$) between modes of consecutive order and same degree is constant and can be approximated as

\begin{equation}
\Delta P=2\pi\,\left(L\,\int_{x_0}^{1}{\frac{N}{x} {\rm d}x}\right)^{-1}\;{\rm ,}
\label{eq:asy}
\end{equation}

\noindent
where $L=[\ell(\ell+1)]^{1/2}$ (with $\ell$ the mode degree), $x$ the normalized radius, $x_0$ corresponds to the boundary of the convective core and $N$ the Brunt-V\"ais\"al\"a frequency.
Moreover, as shown by \citet{miglio2008}, deviations from a uniform period spacing are very sensitive probes of the chemical composition gradient in the star that develops near the edge of the convective core. A sound detection of period spacing in SPB stars would therefore allow direct inferences on the near-core structure of B stars. A theoretical estimate of the expected average $\Delta P$ for $\ell=1$ modes is reported in Fig. \ref{fig:deltap}, where main-sequence models between 2.5 and 8 $M_\odot$, computed with and without overshooting, are considered. We have to recall, however, that such a regular period spacing could be strongly affected by the effects of rotation on such long oscillation periods (see e.g. \citealt{dziembowski1993b} for a discussion).

Analogously to the case of g-mode periods, a regular frequency spacing can occur for p modes. Other situations where a form of spacing can be found, is in frequency multiplets originating from stellar rotation, where deviations from the equidistance case can also occur. A systematic search for recurrent frequency/period spacings therefore represents a valuable tool for a theoretical interpretation of the large number and variety of pulsating stars detected by CoRoT. The total time span of 55 days for the Initial Run puts an upper limit of $\sim 0.02$\,d$^{-1}$ on the frequency resolution. For rich SPB pulsators, the frequencies are expected to be densely packed, thus detecting period spacings may be difficult.

To detect spacings in a list of frequencies $(f_1,\ldots,f_n)$ or periods
$(P_1,\ldots,P_n)$, we use the Kolmogorov-Smirnov test of \citet{kawaler1988} and the inverse variance test of \citet{odonoghue1994}. In addition, a third new test is proposed here.
For the latter, we first construct a vector containing all possible differences between the
periods. This way, we are left with a vector of length $\mbox{Binom}(n,2)$, and
elements $\Delta_{ij}=|P_i-P_j|$. For a test spacing $\Delta P$, we
calculate for each element of these entries $\Delta_{ij}$,
\begin{equation}r = \Delta_{ij} - 
\left\lfloor\frac{\Delta_{ij}}{\Delta P}\right\rfloor\Delta P,
\label{eq:matrix_test}\end{equation}
where $\lfloor x\rfloor$ is the greatest integer smaller than $x$. We count
the number $s_o$ of elements in the intervals
\[s_o \rightarrow [0,\epsilon]\cup[\Delta P-\epsilon,\Delta P]\cdot\]
This number $s$ captures the observed extreme values, which are defined through
the customizable tolerance parameter $\epsilon$. If we assume that these
differences are uniformly distributed, we can express the expected number $s_e$
of extreme values by
\[s_e = \frac{2\epsilon}{\Delta P} n_i.\]
Application of the plain matrix test implies computation of the $p$ value
according to
\[ P(X_i\in[0,\epsilon]\cup[\Delta P-\epsilon,\Delta P]) = 
\frac{\epsilon}{\Delta P}.\] 
This test is appropriate if some periods are missing or
not detected. However, we may prefer to only take those pairs into account for
which the difference between the two members is \emph{exactly} $\Delta P$, and
not a multiple. This we can do using the distinct number test. It works the same
as the matrix test, except that we keep track of the term
\[m=\left\lfloor\frac{\Delta_{ij}}{\Delta P}\right\rfloor\Delta P\]
from Eq.\,(\ref{eq:matrix_test}). Only those spacings where $m=1$ and
$r<\epsilon$ or where $m=0$ and $\Delta P-r<\epsilon$ are considered as valid
period-spacing candidates. The possible disadvantage of using this test is that
we can end up with a lot of periods belonging to the same $\Delta P$, but where
no single triplet is detected. Of course, this can be an advantage as well,
e.g., in the search for rotational splittings; the matrix distinct number test will
detect doublets with the same spacing, far apart around different
frequencies. Finally, if we are interested in the number of periods belonging to
each test spacing $\Delta P$, we can simply count the number of different
periods belonging to each test spacing period set.

\begin{figure}
\includegraphics[width=\columnwidth]{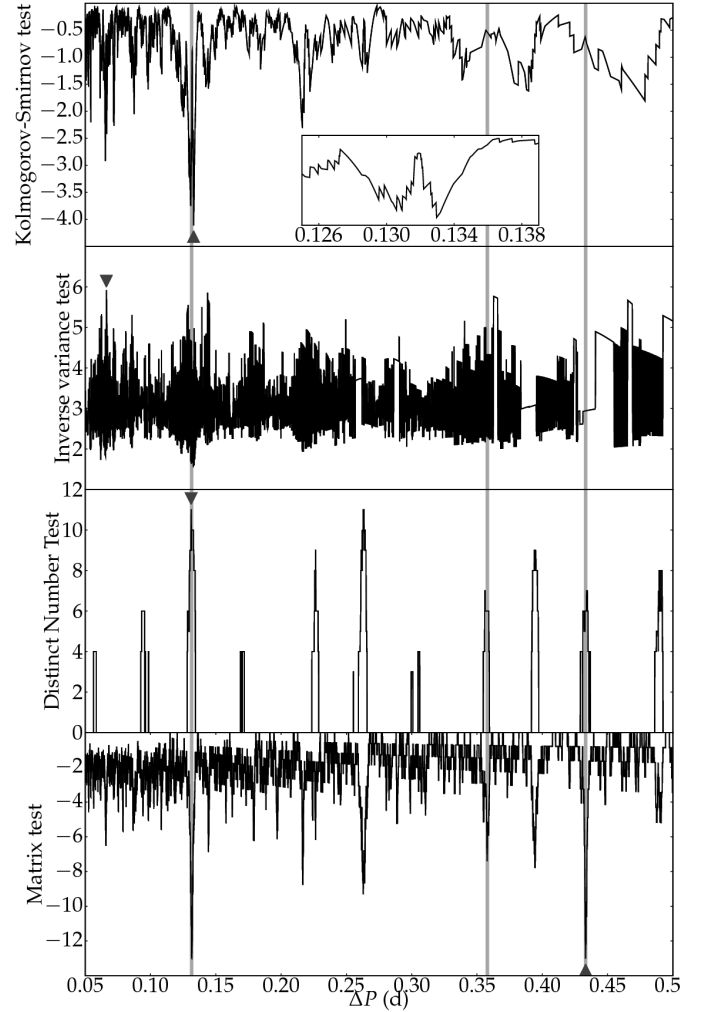}
\caption{Outcome of different tests for period spacings on a simulated set of periods. Input spacings $\Delta P_i$ and detected spacings $\Delta P'_i$ were equal for $\Delta P_1=0.1315$ and $\Delta P_2=0.4332$. For the third spacing $\Delta P_3=0.3579$, the detected spacing was $\Delta P_3=0.3576$.  The (\emph{top panel}) Kolmogorov-Smirnov test with the typical bifurcation, and inverse variance test (\emph{second panel}) do not detect all spacings. The distinct number test (\emph{third panel}) and matrix test (\emph{bottom panel}) are more suitable to detecting the spacings. }\label{fig:spacings}
\end{figure}

To illustrate the performance of these tests, we simulate a
collection of periods, in which we have put 3 different spacings (in arbitrary
units): $\Delta P_1=0.1315$ (8 periods), $\Delta P_2=0.4332$ (6 periods), and
$\Delta P_3=0.3579$ (3 periods). On top of that, we added 8 random periods, to
arrive at a total number of 25 periods. We additionally allowed each period to deviate from the rigid spacing according to a
random normal fluctuation of $\sigma=0.0005$. The results of the different tests
are shown in Fig.\,\ref{fig:spacings}: the Kolmogorov-Smirnov test locates only
the most obvious period spacing, but suffers from a strong bifurcation artifact,
already noticed by \citet{kawaler1988}. The inverse variance test shows lack of
power in this dense pack of periods; it is only able to detect a spacing of
$\Delta P_1/2$. This contrasts to the matrix tests; their
statistic is almost devoid of noise, and only show peaks at the true period
spacings and their multiples. The matrix test and
distinct number test show great similarity in their peak structure,
because they have the same basic calculation method. The reason the matrix
test favours the larger spacing $\Delta P_2$, although it has fewer members than
$\Delta P_1$, is how we set up the statistic: we gave larger spacings
less chance of happening.

Regardless of the succes of these tests, we must be careful in interpreting the
results: because of the occurrence of other random periods, we may by chance
pick up other pairs of periods with the same spacing, but totally unrelated.

\section{Evaluation of the new pulsators}
\subsection{Additional Str\"omgren photometry}
To determine the effective temperature $T_{\rm eff}$ and gravity $\log g$ of the
stars in the initial run, Str\"omgren photometry was obtained using the Wide
Field Camera attached to the Isaac Newton Telescope at the La Palma observatory,
Spain. Because of time constraints, no $\beta$ index measurements could be
obtained. This implied that we did not have a direct measurement of the
gravity. We thus proceeded in an iterative way. First, we computed the
dereddened photometric indices by using an appropriate spectral type for each of
the pulsators: B2V for the candidate $\beta\,$Cep stars, B5V for the
candidate SPBs, A2V for the $\delta\,$Sct candidates, and F2V for the
$\gamma\,$Dor candidates. Next, we used $c_0$ to predict the $\beta$ index
following the relation between these two quantities derived by
\citet{balona1994} for main sequence stars. In fact, \citet{balona1994} has
shown that this procedure has advantages over using the measured indices to
compute $\beta$, due to imperfections in the calibrations. With these dereddened
indices and $\beta$ estimates, the effective temperatures and gravities were
subsequently estimated, using Balona's~(1994) calibration.

The procedure described above delivers an estimate of $T_{\rm eff}$ and $\log g$
with an internal error, i.e., an error assuming that the dereddened input
colours are error-free. It is well known, however, that this procedure for
determining the fundamental parameters suffers from systematic uncertainties
connected with the limitations of the model grids used to explain the behaviour
of standard stars. To get a handle on these systematic errors, we also
used the calibrations of \citet{moon1985}, which are valid along the entire
main sequence as well. The standard deviation between the values for $T_{\rm eff}$ and
$\log g$ derived from Balona's~(1994) and Moon \& Dworetsky's~(1985)
calibrations was considered to be a good approximation of the systematic
uncertainties, so these were added to the statistical errors.

\begin{figure}
\centering\includegraphics[width=\columnwidth]{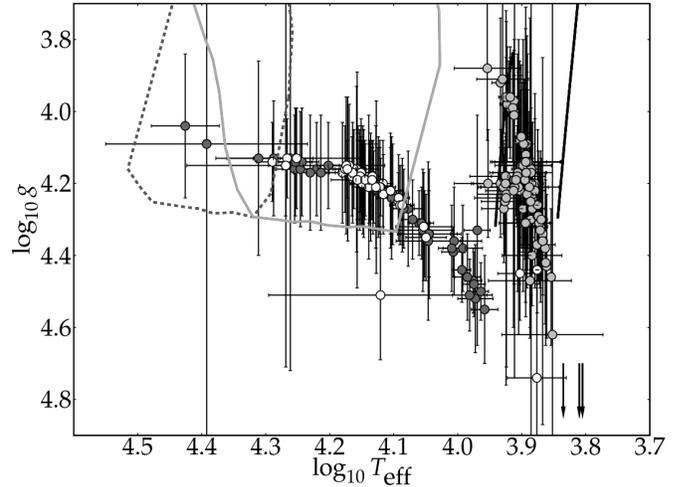}
\caption{Location of targets in a \tgdiagram. The SPB star candidates (\emph{white dots}, 66 stars) occupy the theoretical instability strips well (\emph{grey dashed lines}), except for a few
outliers (see text). The same holds for the 92 $\delta$\,Sct candidates
(\emph{light grey dots}), but not for the 43 $\beta$\,Cep candidates
(\emph{dark grey dots}). Arrows denote outliers: the cool outlier CoRoT
102818325 from the $\beta$\,Cep candidates is probably a binary. The two cool
outliers on the $\delta$\,Sct side each have only two detected frequencies:
CoRoT 102921009 has $f_1\approx 8$\,d$^{-1}$,$f_2=f_1/2$ (possible a binary),
and CoRoT 102787451 has $f_1\approx 17.4$\,d$^{-1}$,$f_2\approx
16.5$\,d$^{-1}$.}\label{fig:sample}
\end{figure}

As a test of the validity of the obtained fundamental parameters for the new
B-type candidate pulsators, we placed all stars in the HRD and compared their
position with the instability strips (Fig.\,\ref{fig:sample}). The numerous new
$\delta\,$\,Sct stars match the classical instability strip perfectly, which
gives us confidence that the estimates of $T_{\rm eff}$ and $\log g$ obtained
from our procedure are appropriate.

From Fig.\,\ref{fig:sample} we see that the classification
is very powerful for extracting candidate new pulsators from a large
sample on the basis of white-light photometry alone (as done in \citet{debosscher2009}), but it is not perfectly reliable on the scale of
individual stars. Moreover, we do not expect so many candidate
$\beta\,$Cep stars compared to SPBs and $\delta\,$Sct stars, considering the mass-dependence of the initial
mass function and of main-sequence lifetimes: this implies that several of them might have a different
character. 
We therefore decided to regroup all new candidate B stars in the sample, by using the 
$T_{\rm eff}$ and $\log g$ information, along with the obtained frequency spectra.
This additional classification should allow a better
distinction between the classes of pulsators along the main sequence than just
using the white-light photometry alone, since we add information on the
fundamental parameters. The new clusters we obtain in this way are discussed by
means of typical and atypical examples, some of which are shown in the text, while others are reported in the electronic Appendix A.

\subsection{Candidate pulsators situated between the SPB- and $\delta$\,Sct instability strips\label{sec:bceps}}

The most obvious result from Fig.\,\ref{fig:sample} is undoubtedly the
appearance of p-mode type variations (in terms of frequency values), not only
far from the theoretical $\beta$\,Cep instability strip, but also outside
\emph{any} instability strip. Part of this discrepancy can be attributed to the CVC not containing
classes of spotted or differentially rotating stars, or ellipsoidal variables. Unavoidably, the main frequency overlaps with the $\beta$\,Cep pulsation range for some of these stars. Moreover, many hot stars in our sample turn out to be good Be star candidates (Neiner, private communication). These are
or will be studied and published elsewhere \citep[][ and future papers]{gutierrezsoto2008,emilio2009}.

The only remaining $\beta$\,Cep candidate is CoRoT 102813271, where two frequencies
are detected in the typical p-mode frequency regime:  $f_1\approx5.79$\,d$^{-1}$, 
$f_2\approx5.51$\,d$^{-1}$ and a third frequency $f_3\approx0.23$\,d$^{-1}$ in the g-mode regime. Because of its
low amplitude and long period, the g-mode frequency can possibly be influenced by instrumental effects.

\begin{figure*}
\centering\includegraphics[width=2\columnwidth]{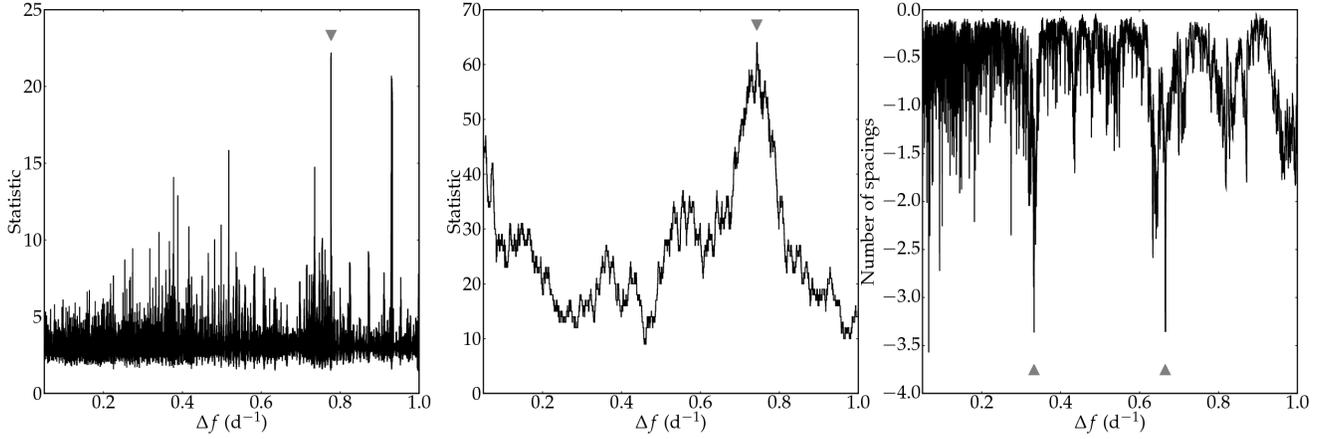}
\caption{Selection of spacing test results: (\emph{left}) the inverse variance
test of detected frequencies in CoRoT 102922479 indicates a spacing of
0.78\,d$^{-1}$. The matrix number (\emph{middle}) test gives a clear spacing of
0.74\,d$^{-1}$ in CoRoT 102848985. The Kolmogorov-Smirnov (\emph{right}) test
applied to CoRoT 102833548 reveals a spacing of $\Delta P_1=0.66$\,d$^{-1}$, but
does not exclude $\Delta P_1/2$ as an alternative
possibility.}\label{fig:spacing_stars}
\end{figure*}

\begin{figure*}
\centering\includegraphics[width=2\columnwidth]{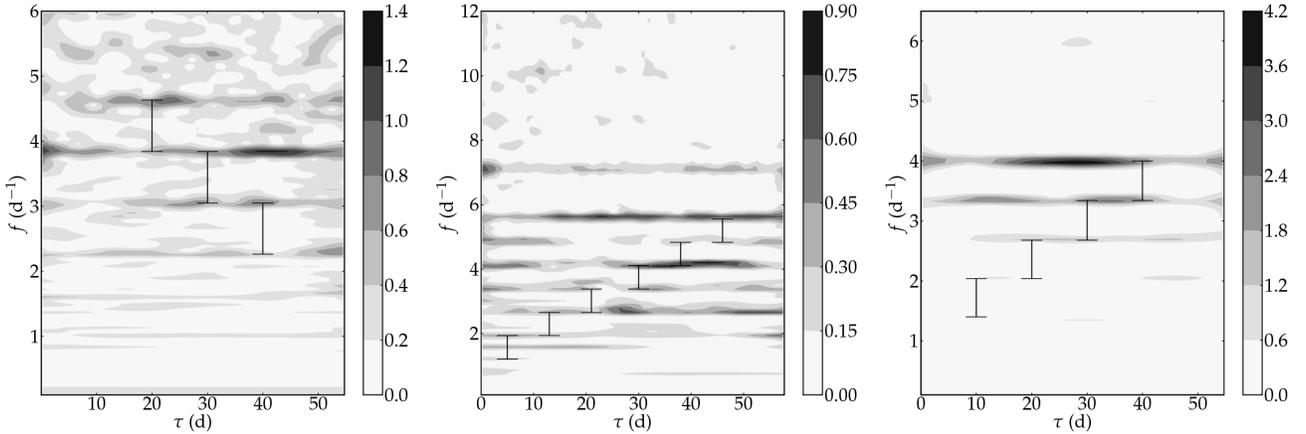}
\caption{Time-frequency analysis of CoRoT 102822479 (\emph{left}), 102848985
(\emph{middle}), and 102833548 (\emph{right}). The spacings found (see
Fig.\,\ref{fig:spacing_stars}) and their derived frequency members are depicted
by the horizontal short lines, vertically connected.}\label{fig:wavelets_stars}
\end{figure*}

The objects on the red side of the instability strips do not constitute one uniform group. That
they all vary over the same timescale implies that the CVC classified them as $\beta$\,Cep stars. However, looking beyond
the first few frequencies reveals their distinct characteristics: CoRoT 102833548, 102848985, 102922479, and to a lesser extent also the almost constant star CoRoT 102850576,
have a well-structured frequency spectrum consisting of equally spaced frequencies (Fig.\,\ref{fig:app:patterns_freqbar} in the Appendix). The results of the spacing tests 
and a time-frequency analysis of the detrended light curves are shown in
Figs\,\ref{fig:spacing_stars} and \ref{fig:wavelets_stars}, and indicate that a spacing of $\sim0.70\,$d$^{-1}$ is common. The amplitude of the main peaks are of the order of 0.1\%. The frequency range overlaps that of $\beta$\,Cep stars
on the low side, but also reaches lower frequencies. The determined temperatures place all these stars consistently between the SPB and $\delta$ Sct instability strips.

CoRoT 102861067, 102790063, 102790331, 102729531, 102862454 and 102771057
constitute a second well-separated group. They share a complex frequency
spectrum containing between $\sim$20 and $\sim$170 significant peaks spread over
a wide range and amplitudes below 0.1\%. Two examples are shown in
Figs\,\ref{fig:app:rp_BCEP_102729531} and \ref{fig:app:rp_BCEP_102790063},
while schematic oscillation spectra are shown in
Fig.\,\ref{fig:app:complex_freqbar} in the Appendix. Dense `forests' of closely
spaced peaks are identified around some frequencies. Based on the 55 days of
photometric observations alone, it is hard to make the distinction between real
multiplets and non-stationarity of the amplitudes.

Close to the $\delta$\,Sct instability strip, there is another small group of
multiperiodic variables: CoRoT 102933855, 102703484, 102850502, and
102816758. Although these stars have a rich frequency spectrum, they are
different from the previous group because their spectrum shows more isolated
frequencies with higher amplitudes ($\sim0.5\%$). Schematic oscillation
spectra are provided in Fig.\,\ref{fig:app:highampl_freqbar}.

Near the border of the SPB instability strip, four monoperiodic variables are
visible: CoRoT 102921797, 102774512, 102889144, and 102872474. The distinction
between monoperiodic oscillation and spots is difficult to make without colour
or spectral information.

The main power of all of the above stars is located in the expected
$\beta$\,Cep pulsation range. However, some of the stars on the red side of the
SPB instability strip show variability on longer time scales ($<4$\,d$^{-1}$),
e.g. Figs\,\ref{fig:app:erratic_motions_spb} and \ref{fig:app:corot_102808565_highfreq}
in Appendix A. Not all of this periodic variability is sinusoidal or even
approximately sinusoidal. The best example in this respect is CoRoT 102762284,
for which the sinusoidal variation is superposed on a larger scale outburst-like
variation (Fig.\,\ref{fig:app:outburst}).

\begin{figure*}
\centering\includegraphics[width=2\columnwidth]{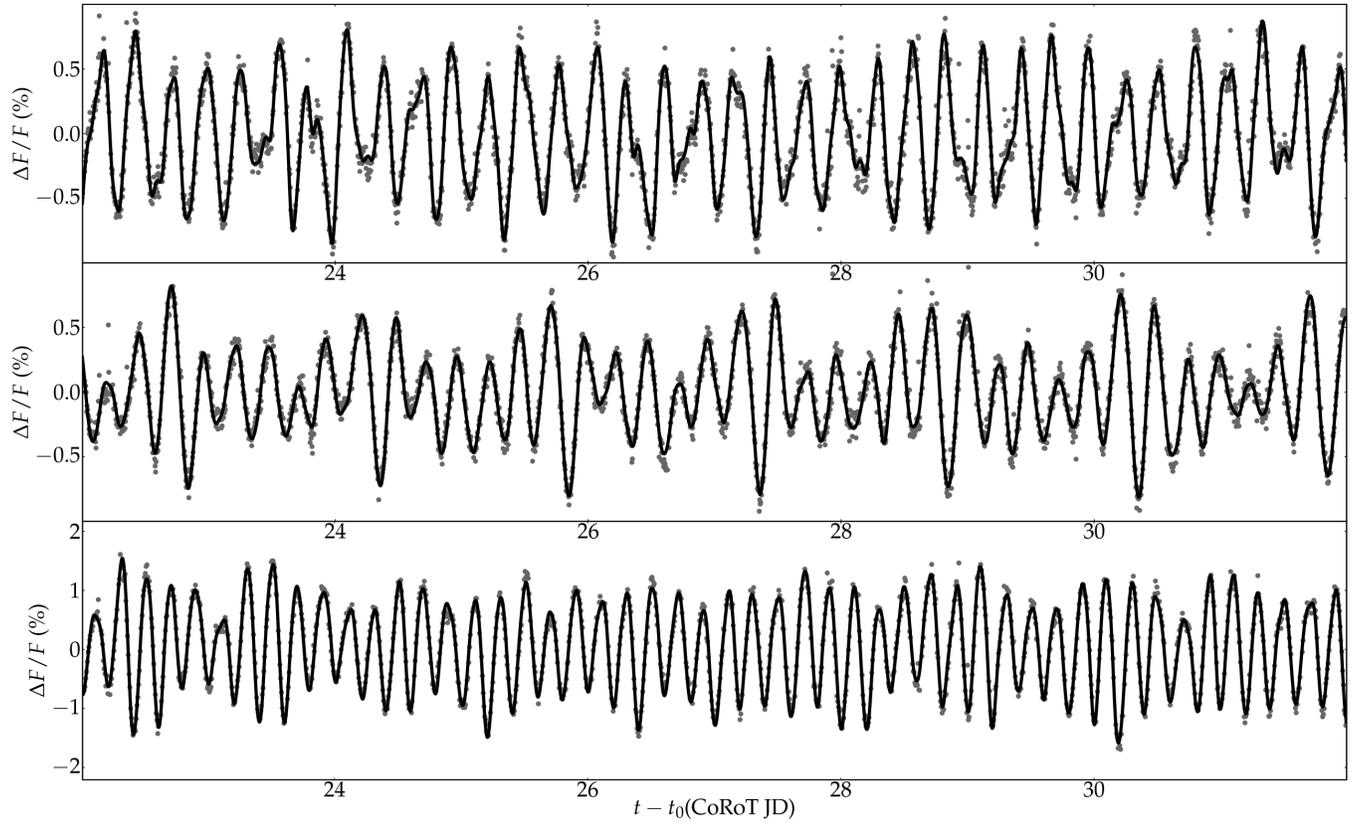}
\caption{B-type pulsator candidates on the red side of the SPB instability
strip. Some of these stars show clear, stable oscillations. \emph{(top)} CoRoT
102816758 ($t_0$=2593.0705 CJD, $f_1\approx3.6$\,d$^{-1}$), \emph{(middle)}
CoRoT 102833548 ($t_0$=2593.0705 CJD, $f_1\approx4.0$\,d$^{-1}$),
\emph{(bottom)} CoRoT 102850502 ($t_0$=2593.0705 CJD,
$f_1\approx5.0$\,d$^{-1}$). Light curves with high time resolution were binned
per $\sim$10 datapoints for visibility reasons. The black line represents the 
fit from which the satellite orbital
frequency, its harmonics, and its aliases 
were filtered out.}\label{fig:clear_motions}
\end{figure*}

\begin{figure}
\centering\includegraphics[width=\columnwidth]{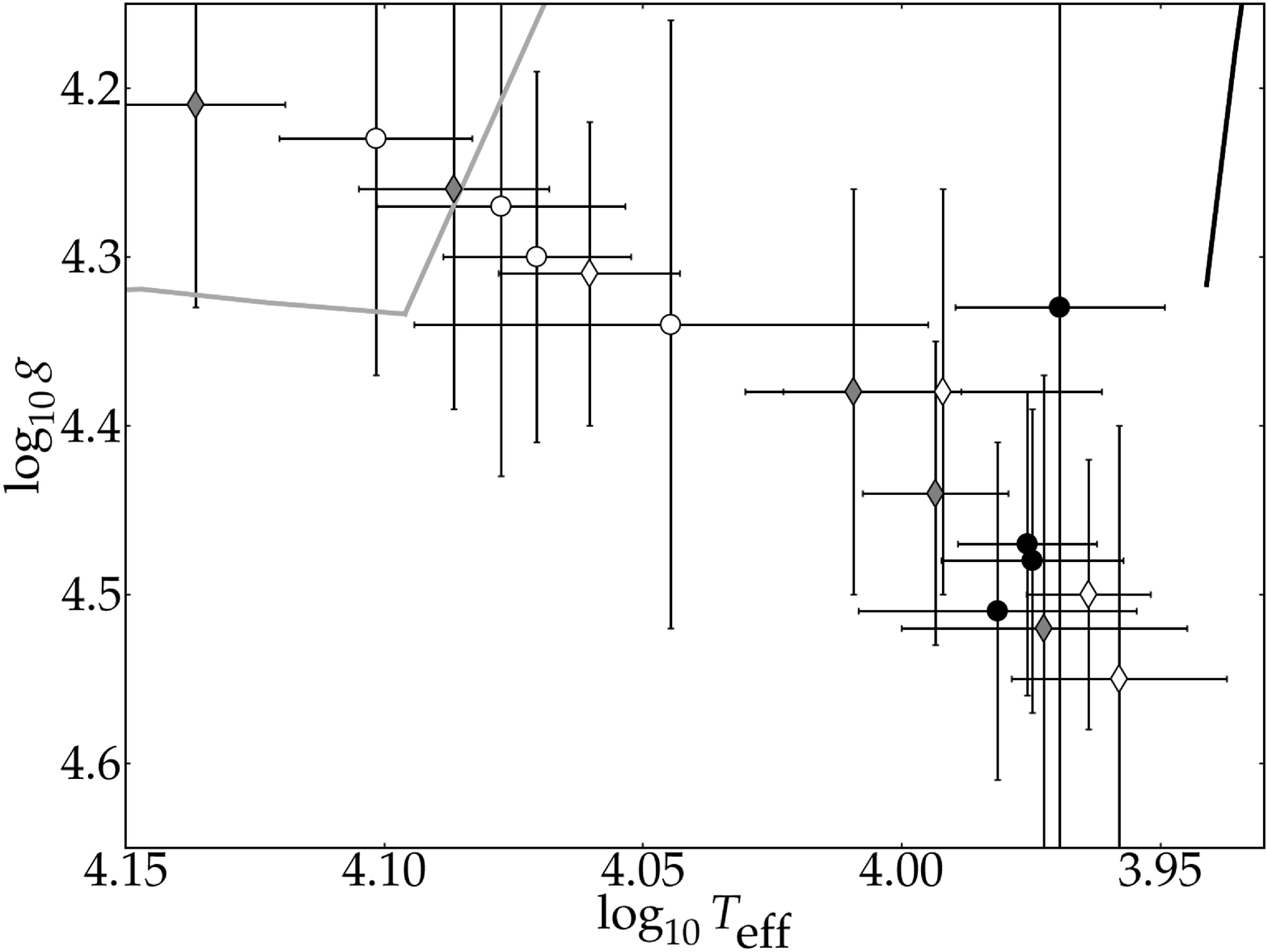}
\caption{Location of $\beta$\,Cep candidate pulsators in the $\log T_{\rm
eff}-\log g$ diagram with a determined $T_{\rm eff}$ too low to be a real
$\beta$\,Cep star: $\delta$ Sct like variables (black dots), monoperiodic
$\beta$\,Cep candidates (white dots), rich pulsators (grey diamonds) ,and targets
with equidistant frequency spacing (black white
diamonds).}\label{fig:bcep_summary}
\end{figure}

A summary of the interpretation of the cool, short-term
(i.e. $\beta$\,Cep-like) candidate pulsators in the $\log T_{\rm eff}-\log g$
diagram is shown in Fig.\,\ref{fig:bcep_summary}.

\begin{figure}
\centering\includegraphics[width=\columnwidth]{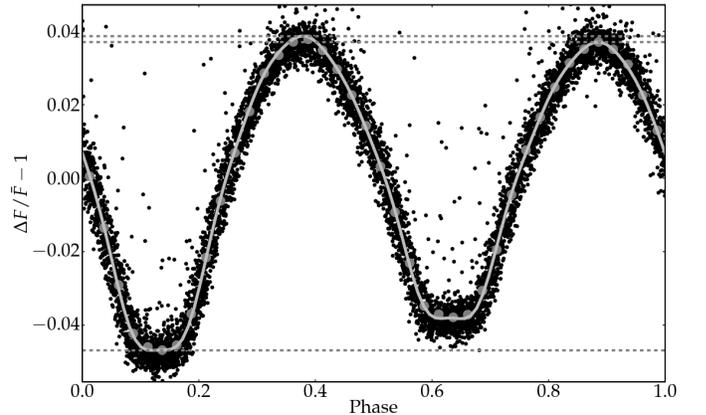}
\caption{Binary system CoRoT 102725806 with an orbital frequency
$f=2.925(7)$\,d$^{-1}$. Black circles are data points, dark grey circles are
phase-binned data points, the light grey line is a fit using 7 harmonics,
determined via consecutive prewhitening.}\label{fig:binary_candidate}
\end{figure}

\subsection{SPB candidate examples}

\begin{figure}
\centering\includegraphics[width=\columnwidth]{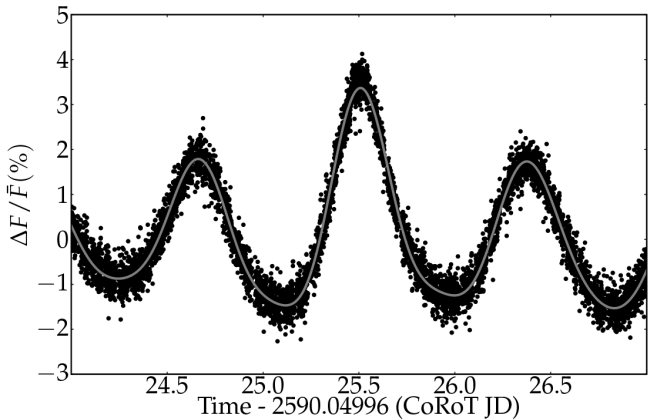}
\caption{CoRoT 102938439: excerpt of original light curve and fit (solid grey
line), around a time where the amplitude is temporarily driven to a higher
level. The typical shape of shallow minima and sharp maxima is
visible.}\label{fig:spb_shape}
\end{figure}

\begin{figure*}
\centering\includegraphics[width=2\columnwidth]{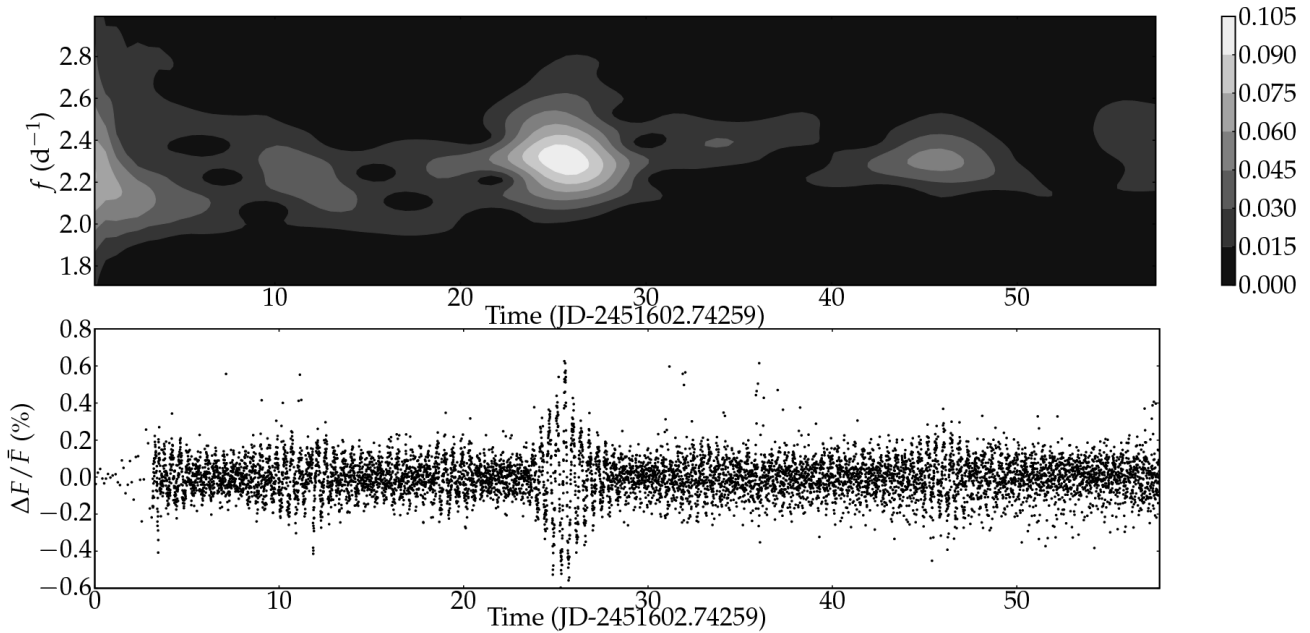}
\caption{(\emph{top panel}) Short Time Fourier Transformation of the residual
light curve of CoRoT 102938439, after prewhitening a nonlinear fit with all
detected significant frequencies below $\sim$1.5\,d$^{-1}$. The highest
amplitude frequencies in the residuals clearly show modulated
amplitudes. (\emph{bottom panel}) A binned version of the residual lightcurve
clearly shows the largest temporal variability near day 25 of the light curve,
with some smaller `outbursts' as well.}\label{fig:corot_102938439}
\end{figure*}

\begin{figure}
\centering\includegraphics[width=\columnwidth]{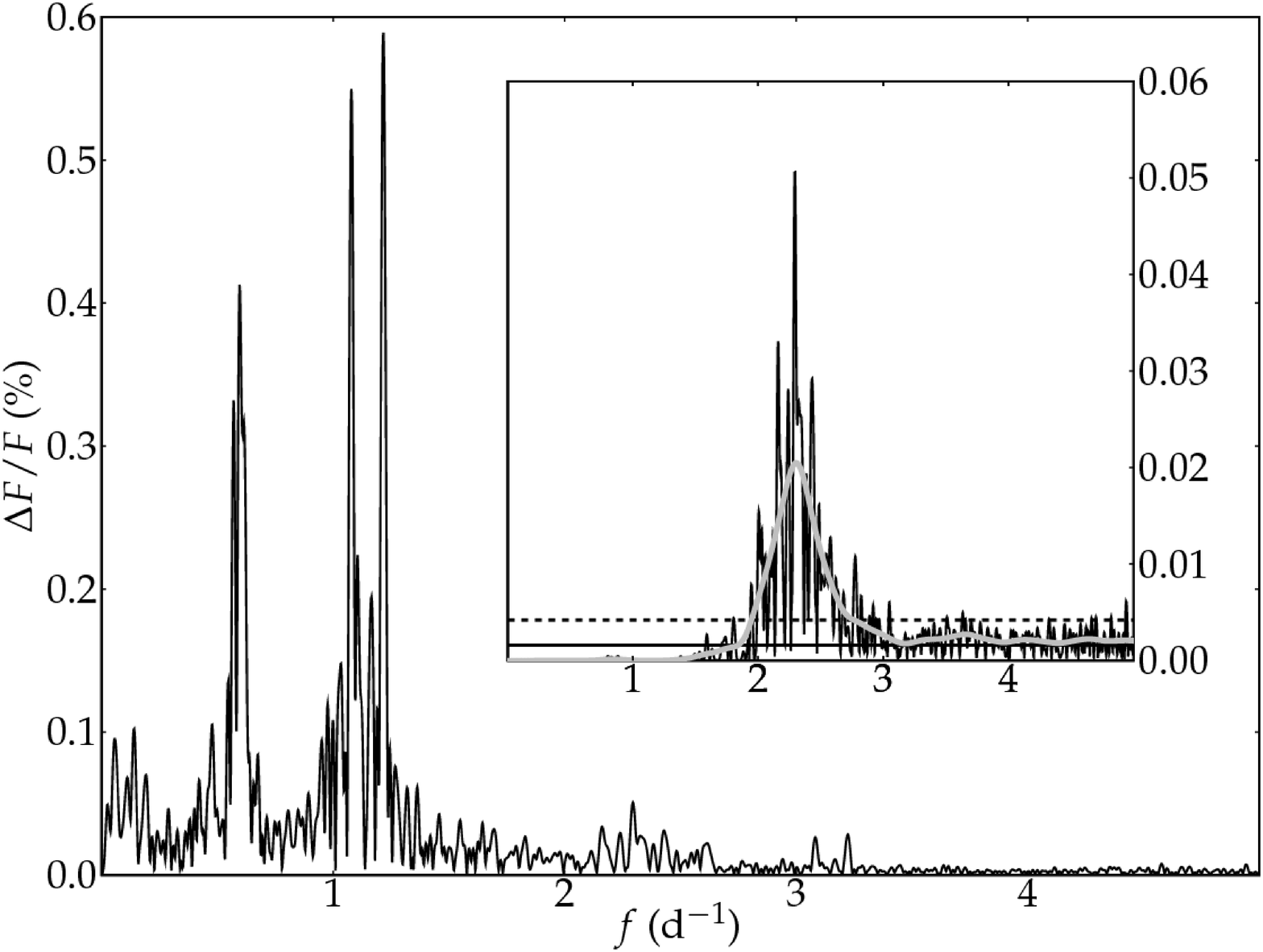}
\caption{Original periodogram of CoRoT 102938439. The inset is the amplitude
spectrum of the residuals, after prewhitening all significant frequencies below
$\sim1.5$\,d$^{-1}$. Grey solid line is a Gaussian smoothing with
$\sigma=0.075$\,d$^{-1}$. The black solid line denotes the mean noise level, the
dotted line is the $99\%$ significance level.}\label{fig:power_excess}
\end{figure}
As is clear from Fig.\,\ref{fig:sample}, the SPB star candidates from the
initial classification agree with their instability strip.  The 15
clearest SPB candidates, in order of decreasing amplitude, from $\sim 1\%$ to
$\sim 0.2\%$, are CoRoT 102844894, 102754851, 102728830, 102855391, 102739246,
102797587, 102938439, 102769848, 102848506, 102848506, 102956197, 102943966,
102968674 and 102871668. Their effective temperature and gravity locates them
well within the theoretical SPB instability strip. Besides the typical closely
spaced g mode pulsations, we also find a residual power excess bump in a
frequency band above 1.5-2\,d$^{-1}$ for the majority of these stars, well
separated from the typical SPB pulsation range predicted for excited modes of
$\ell<3$ modes. The theoretical instability strip reaches the observed range of
frequencies for modes with a higher $\ell$ value, but at the same time, disk-averaging
effects reduce the observed amplitudes. For $\ell=4$, this reduction
factor amounts to about 30, which is still above the CoRoT noise level for many
of the SPB candidates, so we cannot exclude 
the higher observed frequencies being due to high-degree
modes. We illustrate the nature of the observed variability for a
prototype of these stars (CoRoT 102938439) in Figs.\,\ref{fig:corot_102938439},
\ref{fig:power_excess}, and \ref{fig:spb_shape}. Other examples can be found in
Figs.\,\ref{fig:app:power_excess1} to \ref{fig:app:power_excess5}). After
prewhitening all low frequencies, the largest feature in the amplitude spectrum
turns out to be a collection of closely spaced peaks
(Fig.\,\ref{fig:power_excess}). At least two phenomena can cause these
features: unresolved stable modes with a complex beating pattern, or
time-dependent oscillation frequencies and/or amplitudes. In the first case, we
expect a region with blended or nearly blended frequency peaks, without any
amplitude structure. In the second case, we expect a frequency region with a
peak at the excited frequency, and Lorentzian shape farther away from the
central peak. When a residual light curve of these stars is constructed by
prewhitening a nonlinear least squares fit of all frequencies below
$\sim$1.5\,d$^{-1}$, the same amplitude structure emerges for all of them,
which points towards the same physical origin given that the pattern is not
chaotic.  The
amplitudes of the detected modes are highly variable over time (see
Figs\,\ref{fig:app:power_excess1} and \ref{fig:app:power_excess2}). The observed
features could come from modes with a finite lifetime
or to nonlinear nonresonant distortion leading to
time-dependent phenomena. An analysis of the evolution of the
amplitudes shows that the period over which the modes appear and dissappear is
of the order of one to a few days. 
Instead of a sinusoidal shape, the light curve maxima are
narrower and sharper, while the minima are shallow and broad
(Fig.\,\ref{fig:spb_shape}). This is very similar to the light curves of some
white dwarf pulsators and is explained by nonlinear effects for these objects
\citep{vuille2000}. We notice that a similar feature of disappearing and
reappearing frequencies was also found in the Be star HD 49330 and was
interpreted as the cause of an outburst \citep{huat2009}. It could very well be
that this is a general phenomenon at low amplitudes for B stars, but that it only
leads to mass loss for the most rapid rotators among them.

A second manifestation of nonlinearity can be found in many detected
combination frequencies. In contrast to the time-dependent phenomenon described
above, these are stable on the scale of the total time span, thus resulting in
isolated frequency peaks in the spectra (see online frequency tables). A
detailed description of these modes is difficult based on the short time span of
the Initial Run and the long periods of these modes ($<2$\,d$^{-1}$), but will
surely be possible for similar new pulsators in the CoRoT long run data.

Secondary good SPB star candidates include CoRoT 102863407, 102752912,
102773435, 102917802, 102826973, 102838201, 102888003, 102915048, 102813396, and
102945383 of which the last four also show evidence for modes of finite
lifetime.

\subsection{Other variable objects}

Brightness variations need not necessarily be due to pulsations
(alone). For example, when two binaries are physically close to each other and
the stars are of comparable luminosity and size, and the grazing eclipses 
can resemble one
single, stable sinusoidal pulsation. A DFT analysis puts more power in the
double frequency of the orbital frequency, folding both the primary and
secondary eclipse onto each other. Therefore, close binaries with an orbital
period roughly between 0.25 and 0.5 $d^{-1}$ can easily be mistaken for a
$\beta$\,Cep star, whereas a wider system can be misinterpreted as an
SPB. Examples of close or contact binaries are CoRoT 102725806, 102858055,
102876625, 102930503, 102784048, 102768686, 102732139, 102873761, and
102812372. The phase diagram of the first candidate, CoRoT 102725806, is shown
in Fig.\,\ref{fig:binary_candidate}. We find an orbital frequency of
$f=2.925(7)$\,d$^{-1}$, independently confirmed with the phase dispersion
minimization and information entropy method. The amplitude of the primary
eclipse is $A_1=8.56\%$, compared to the average flux level. The double system
was mistaken for a single $\beta$\,Cep star by the CVC because of the small
difference between the amplitude of the primary and secondary eclipse;
$A_1-A_2=0.87\%$. The duration of the primary eclipse in phase units is
$P_1\approx 0.08$, which is almost equal to the duration of the secondary
eclipse, $P_2\approx 0.10$.

The possibility of a binary system and an SPB candidate are not mutually
exclusive, because CoRoT 102918586, 102912741 and CoRoT 102793963 turn out to be
candidate SPBs in close binaries, of which the latter shows a single sinusoidal
variation, synchronised with the orbit of the binary. These systems in
particular can be of interest for asteroseismology, because the binary orbit can
help to constrain the mass and radius of the SPB star.

\begin{figure*}
\centering\includegraphics[width=2\columnwidth]{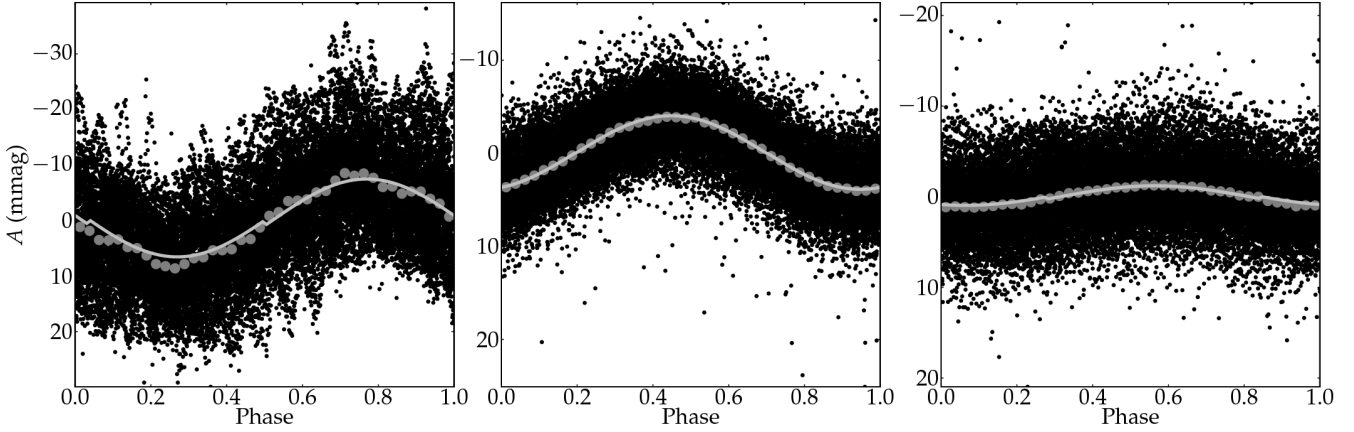}
\caption{Phase diagrams of three clear sinusoidal variations of CoRoT
  102804522. The frequency in the left panel ($f\approx1.04$\,d$^{-1}$) is
  rather high for typical $g$ modes in SPBs, while the frequency in the middle
  panel ($f\approx15.27$\,d$^{-1}$) is rather high for a typical $\beta$\,Cep
  star $p$-mode, but quite typical for $p$ modes in $\delta$\,Sct stars. The
  frequency with which the light curve is folded in the right panel
  ($f\approx4.05$\,d$^{-1}$) could, on the other hand, be a typical $p$ mode in
  a $\beta$\,Cep star. This different behaviour raises the suspicion that this
  is not one single star, although there are neither traces of eclipses nor any 
indication of flux
  contamination from nearby stars on the CCD.}\label{fig:corot_102804522_phases}
\end{figure*}

Another group of SPB candidates have frequencies in a very broad range,
e.g. CoRoT 10282445, 102764403, 102818535, 102816987, 102837646, 102804931,
102887852, 102930369, 102790135, and 102856178. CoRoT 10282445, for example,
shows periodic variations with time scales around $\sim$\,$1$\,d$^{-1}$,
$\sim$\,$4$\,d$^{-1}$ and $\sim$\,$15$\,d$^{-1}$
(Fig.\,\ref{fig:corot_102804522_phases}). The main frequency
$f_1=1.0411(7)$\,d$^{-1}$ can be connected to the third frequency
$f_3=1.1533(3)$\,d$^{-1}$, resulting in a beating period of almost 10 days. The
second frequency $f_2=15.271(9)$ is connected to
$f_{29}=14.739(4)$\,d$^{-1}$, and the same spacing of 0.52\,d$^{-1}$ is found
throughout the whole frequency spectrum
(Table\,\ref{tbl:spacing:102804522}). Although the spacing connects the three
main frequency regions where variability occurs, current stellar models do not
offer any interpretation in terms of a single SPB, $\beta$\,Cep,  or $\delta$\,Sct
star, while we can exclude contamination from nearby stars on the CCD. The
possibility of a binary system consisting of a hybrid $\beta$\,Cep and a
$\delta$\,Sct remains, as well as the one of a triple system. Spectra will help
us to unravel the exact cause of the variation in this light curve.

\begin{figure*}
\centering\includegraphics[width=2\columnwidth]{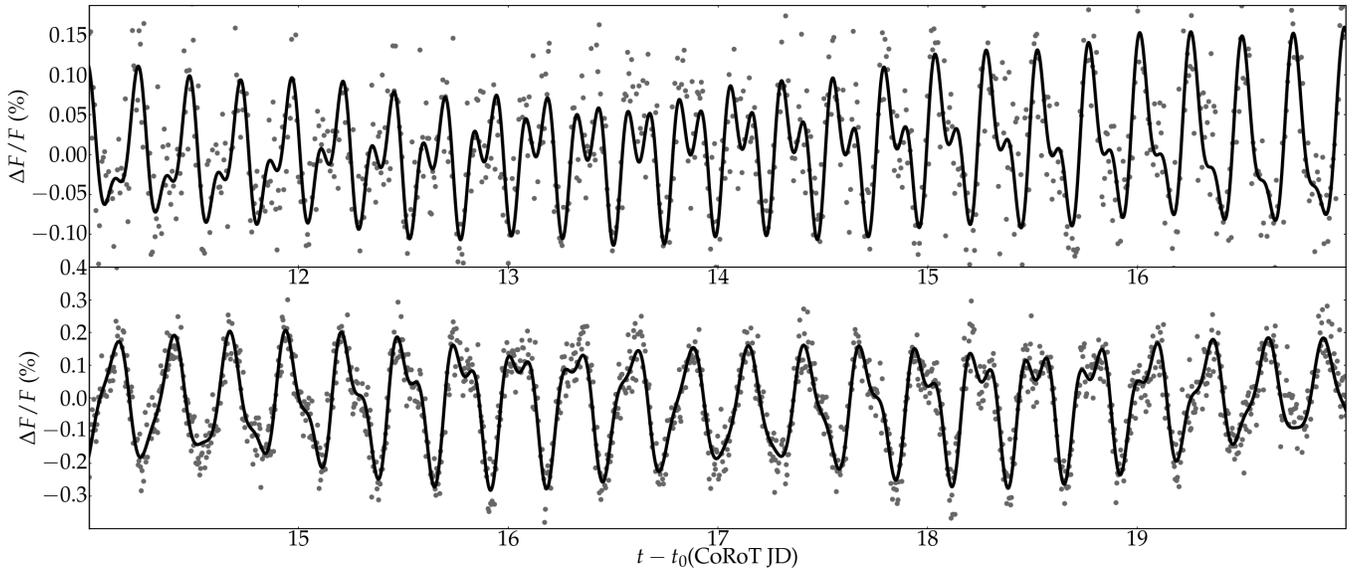}
\caption{Selection of $\beta$\,Cep candidates showing strong hints of spot
features in their light curves. (\emph{upper panel}) CoRoT 102968526
($t_0$=2590.0450 CJD) has a dominant frequency of $f_1\approx4.14$\,d$^{-1}$ with a secondary peak at $f_2\approx4.08$\,d$^{-1}$.
The
black line is a fit using all noninstrumental frequencies (\emph{bottom panel})
The light curve of CoRoT 102857565 ($t_0$=2590.0499 CJD) was binned per 10
points for visibility reasons.}\label{fig:spotted_star}
\end{figure*}

\begin{figure*}
\centering\includegraphics[width=2\columnwidth]{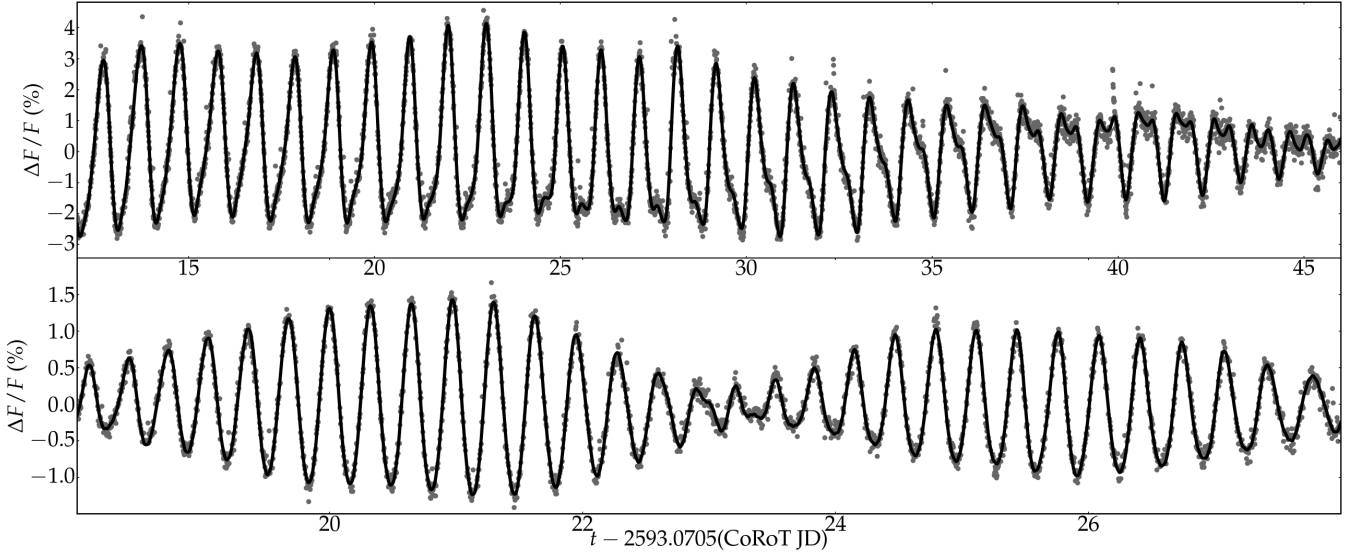}
\caption{CoRoT 102835515 (\emph{upper panel}) and 102824452 (\emph{lower panel},
light curve binned per 10 points) have a frequency spectrum where the power is
concentrated in narrow bands at multiples of the dominant frequency. Combined
with the peculiar shape of their light curves, they strongly resemble spotted
stars, although the distinction from a simple beating pattern is not
easy (\emph{bottom}). The solid line is a fit, grey circles are
data.}\label{fig:spotted_spb_stars}
\end{figure*}

For a couple of light curves (CoRoT 102910610, 102968526, 102857565, and
102876625, see Fig.\,\ref{fig:spotted_star}), 
spot models seem more appropriate for describing them than a pulsation model; 
the resemblance to simulated light curves of
rotating spotted stars is indeed apparent
\cite[e.g.][]{strassmeier1992,lanza1993}. Their frequency spectra 
agree with those models, and few frequencies are detected in a very narrow
interval. Figure\,\ref{fig:spotted_spb_stars} illustrates that it can be difficult to distinguish simple beating patterns from spot related variability.
By \emph{assuming} that the star rotates differentially like the Sun, the
multiplet structure can be used to determine the level of differential rotation
$\Delta\Omega$, e.g., for CoRoT 102968526,
$\Delta\Omega=0.0538\pm0.0002$\,d$^{-1}$. This implies that one part of the
star completes one more full rotation cycle over a period of $1/\Delta\Omega$.

\section{Comparison with theoretical predictions}
 \begin{figure}
\centering\includegraphics[width=\columnwidth]{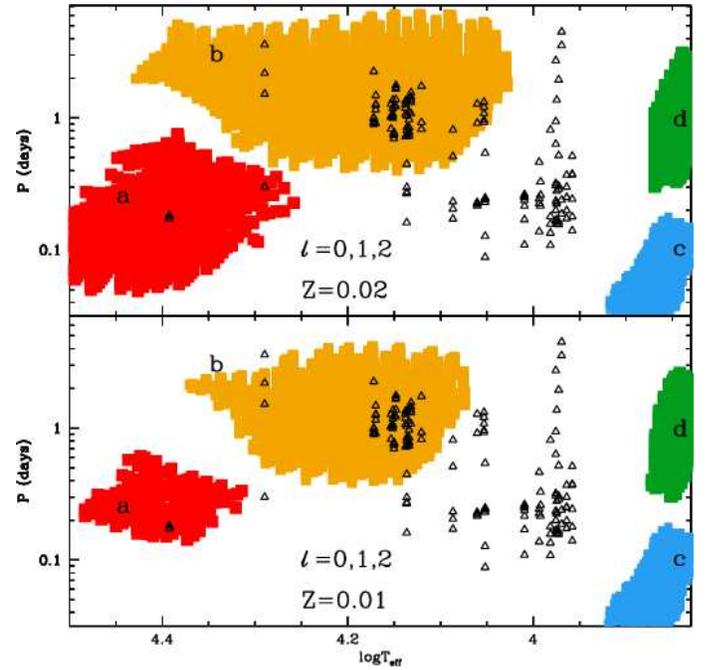}
 \caption{Theoretical instability domains of $\ell=0,1,2$ modes represented in a
 $\log{\teff}$-$\log{P}$ diagram for main-sequence models with $Z=0.02$ (upper
 panel) and $Z=0.01$ (lower panel). In each panel, the regions of unstable modes
 represent $\beta$\,Cep (a), SPB (b), $\delta$ Sct (c), and $\gamma$ Dor-type
 pulsations (d).The frequencies detected in the best SPB candidates (see
 text) and the candidate pulsators from Sect.\,\ref{sec:bceps} are shown as
 triangles.}\label{fig:spb}
 \end{figure}

The results presented in the previous section show that many of the stars
classified as an SPB or $\beta$\,Cep star by the CVC classifier are not typical
members of the two classes (at least as far as the classes have been defined so far).
To have a general comparison between observed and theoretically
predicted instability domains, we not only consider $\log{T_{\rm eff}}-\log{g}$
planes (see Fig. \ref{fig:sample}), but also confront predicted and observed
frequency instability domains, as presented in Fig.\,\ref{fig:spb}.

The theoretical instability domains were computed for nonrotating main-sequence
models of masses between 1.2 and 18 $M_\odot$ and 2 values of the heavy-elements
mass fraction: Z=0.02 and Z=0.01 (respectively upper and lower panels of
Fig. \ref{fig:spb}). Stellar models were computed with {\sc cles}
\citep{scuflaire2008}, adiabatic frequencies with {\sc losc}
\citep{scuflaire2008b}, and the stability of modes of degree $\ell=0,1,2$ was
investigated with the code {\sc mad} \citep{dupret2003}. In the case of $\delta$
Sct and $\gamma$ Doradus stars, the nonadiabatic computations included the
interaction between convection and pulsation as described by
\citet{grigahcene2005}. All stellar models were computed with overshooting from
the convective core ($\alpha_{\rm OV}=0.2$) and a mixing-length parameter
$\alpha_{\rm MLT}=2.0$, both expressed in local-pressure, scale-height units. OP
opacity tables \citep{badnell2005} with \citet{asplund2005} metal mixture were
adopted. We refer to \citet{miglio2007,miglio2007a} and \citet{zdravkov2008} for
a detailed study of the effect of considering different opacity tables and metal
mixtures on the excitation of pulsation modes in B-type stars.  The distinct
islands of instability populating the $\log{T_{\rm eff}}$-Period diagram of
Figs\,\ref{fig:spb} correspond (from left to right) to $\beta$\,Cep-, SPB-,
$\delta$ Sct-, and $\gamma$ Dor-type pulsations.

Stars classified as SPBs are in overall agreement with the theoretically
expected domain (see Figs. \ref{fig:sample} and \ref{fig:spb}), though
in some cases with frequencies higher than those of a typical SPB star (which
could be due to the strong effects of rotation on the periods of oscillation). On
the other hand, several stars classified as $\beta$\,Cep are in the typical
$T_{\rm eff}$ instability domain of SPBs. These stars have frequency peaks in
the $\beta$\,Cep frequency domain: their modes could be prograde SPB-type modes
as proposed in, e.g., \citet{saio2007}.

The effects of rotation may also affect the determination of $T_{\rm eff}$ for a
few stars that have a frequency spectrum resembling the one of fast-rotating
SPBs but that, based on their position in Fig.\,\ref{fig:spb}, are cooler than
the red edge of the SPB instability strip (e.g. CoRoT 102833548 and
102729531). A more reliable spectroscopic $T_{\rm eff}$ determination of these
stars is thus needed to clarify the nature of these targets. Such spectroscopic
data will become available in the next months.

The comparison with theoretical expectations will also greatly benefit from the
determination of the number of pulsating B stars relative to all the B stars
observed in CoRoT's IRa01 EXOfield. The absence of typical $\beta$\,Cep
pulsators (in terms at the same time of frequency spectrum and $T_{\rm eff}$)
can be ascribed simply to the expected scarcity of early B-type stars in the
field, following the initial mass function and stellar evolution models. This
requires the determination of stellar parameters for a large number of stars and
will be addressed in a future work.

It is clear from these first general remarks that the relevance of CoRoT's
photometric variability survey goes beyond the simple definition of instability
domains of expected and known types of pulsators. Particular theoretical
interest is provided by the unexpected variability detected in several targets
on the red side of the SPB instability strip, a region where several claims of
the detection of pulsations were made (the so-called ``Maia'' stars, see
e.g. \citealt{scholz1998}; \citealt{aerts2005}).  The 
theoretical instability
mechanisms proposed as an explanation 
for these stars do not cover the full area in the Hertzsprung-Russell diagram
where such pulsations have been claimed
(see \citealt{townsend2005} and \citealt{savonije2005}). The
forthcoming spectroscopic observations of those targets will shed new light on
these most interesting targets, and allow theoretical interpretation of the
detected variability.

\section{Conclusions}

We constructed a frequency analysis pipeline and additional post-processing
tools to efficiently analyse pulsators in CoRoT's exoplanet
database. The method also includes a jump correction and detrending algorithm,
as well as an automated search for combination frequencies and spacings among
frequencies or periods. In the particular case of the Initial Run, many
combination frequencies and spacings were found, but the total time span of
$\sim55$\,d was too short to draw firm conclusions on the frequency or period
spacings of g modes and to confront them with theoretical models. There were too
few convincing candidates in the more massive $\beta$\,Cep class to commence an
interpretation in terms of nonlinear resonant coupling. However, our techniques
revealed clear frequency spacings in high-frequency variables on the cool side
of the SPB instability strip.

Most of the stars that seemed good $\beta$~Cep candidates in terms of the values
of the first three frequencies are to be refuted. This was to be expected,
because the number of candidate $\beta\,$Cep stars found by the CoRoT
Variability Classifier \citep{debosscher2009} was anomalously high compared to
the number of candidate SPBs. Only one star on the blue side of the SPB
instability strip matches a typical $\beta$\,Cep frequency spectrum.  Four other
stars were found to be $\delta$\,Sct/$\beta$\,Cep-like, but again were far off
their theoretical location in the $\log T_{\rm eff}-\log g$ diagram.  Instead,
we detected many different types of variable stars among the $\beta\,$Cep
candidate sample.  Several turned out to be low-amplitude pulsators whose
frequencies are spread over a wide range and which are located on the red side
of the SPB instability strip, where no oscillations are predicted by
theory. According to the theoretical computations by \citet{townsend2005} and
\citet{savonije2005}, mid-B type rapidly rotating pulsators with retrograde
mixed g~modes could occur in  a part of this region, exhibiting many of these
properties. However, they could also simply be fast-rotating
classical SPB or $\delta$~Sct stars, or Be pulsators.  Future spectroscopic
observations will allow us to distinguish between these possibilities, by
supplying information on rotational velocities, abundances, and improved $\log
g-\log T_{\rm eff}$ values. Besides these interesting stars, also binary stars
and a few suspectedly differentially rotating stars (with or without additional
pulsations), were detected among the $\beta$~Cep candidates.

Similar conclusions apply to the sample of SPB candidates, although the number
of good candidates is vastly greater for this class than for the $\beta\,$Cep
candidates.  Interestingly, the highest amplitude SPB candidates all show
two distinct signs of nonlinear behaviour: first, clear unexpected
power excess in the region between 2 and 6\,d$^{-1}$, which seems to point to
modes with finite lifetimes. These could originate from nonlinear
nonresonant harmonic distortion, leading to time-dependent amplitudes and/or
frequencies. Second, we identified combination frequencies, which
are stable on the time scale of the measurements. These are likely
to be caused by nonlinear resonant mode locking. On top of that, we also find
some rich and low-amplitude SPB pulsators, with frequencies in a lower range
than the similar stars in the $\beta$~Cep sample.

Our follow-up studies of new pulsating B stars in the CoRoT long runs, which
  typically last 150\,d, as well as future spectroscopy, will be very valuable
  in evaluating the new type of oscillations found in the Initial Run data as
  reported here in more detail.

\begin{acknowledgements}
The research leading to these results received funding from the European
Research Council under the European Community's Seventh Framework Programme
(FP7/2007--2013)/ERC grant agreement n$^\circ$227224 (PROSPERITY), as
well as
from the Research Council of K.U.Leuven grant agreement GOA/2008/04 and from
the Belgian PRODEX Office under contract C90309: CoRoT Data Exploitation.
This article is based on observations made with the INT-WFC operated on the island of \emph{La Palma} by the Isaac Newton Group in the Spanish Observatorio del Roque de los Muchachos of the Instituto de Astrofísica de Canarias. This research has been partly supported by the \emph{AYA2006-15623-C02-02 of the MCINN}. 
\end{acknowledgements}

\bibliographystyle{aa}
\bibliography{11884}

\Online

\begin{appendix}

\section{Additional Figures \& Tables}

\clearpage

\begin{figure}
\includegraphics[width=\columnwidth]{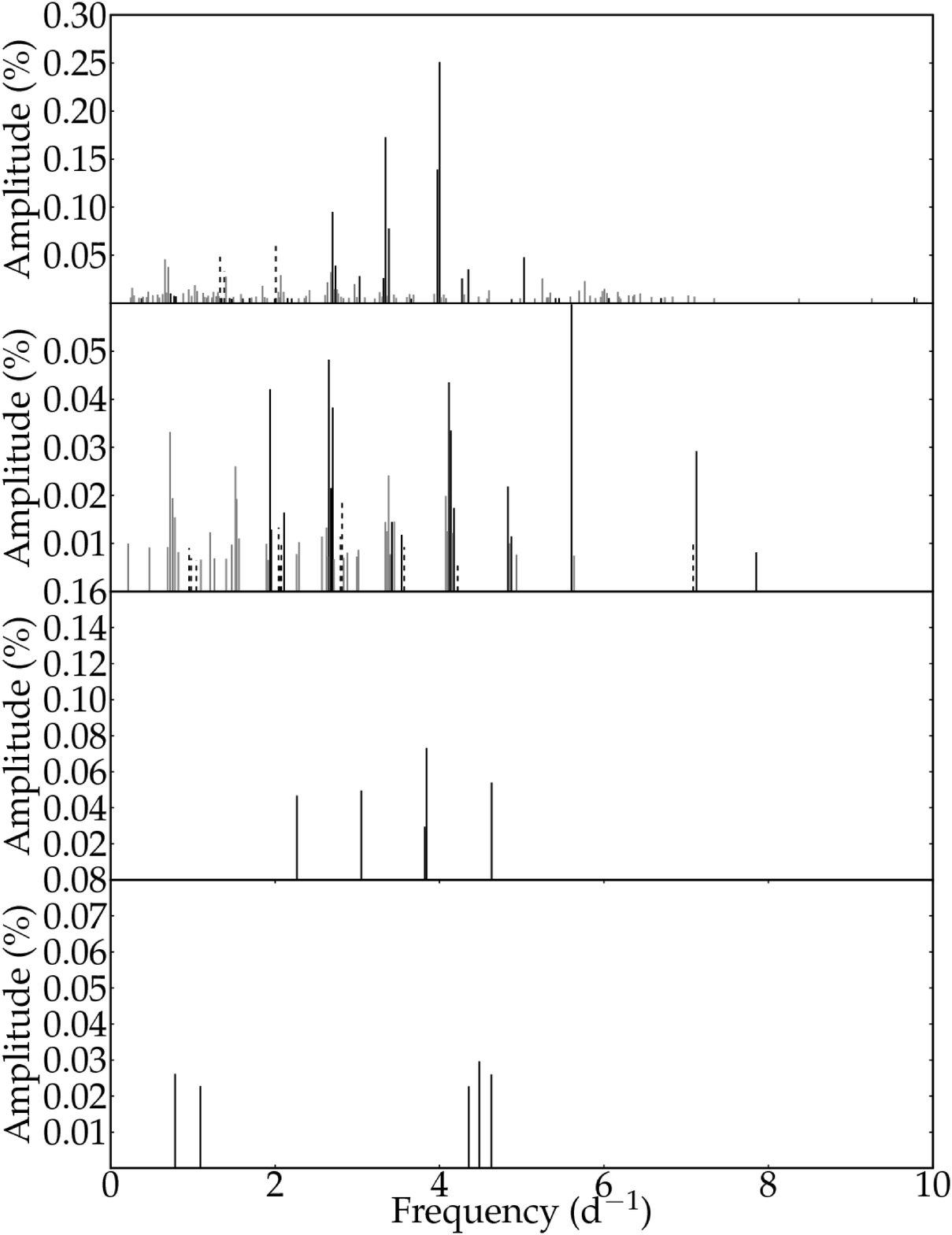}
\caption{Sames as Fig.\,\ref{fig:app:complex_freqbar}, but for the stars showing a complex but structured
frequency spectrum (CoRoT 102833548, 102848985, 102922479, 102850576).\label{fig:app:patterns_freqbar}}
\end{figure}

\begin{figure*}
\centering\includegraphics[width=2\columnwidth]{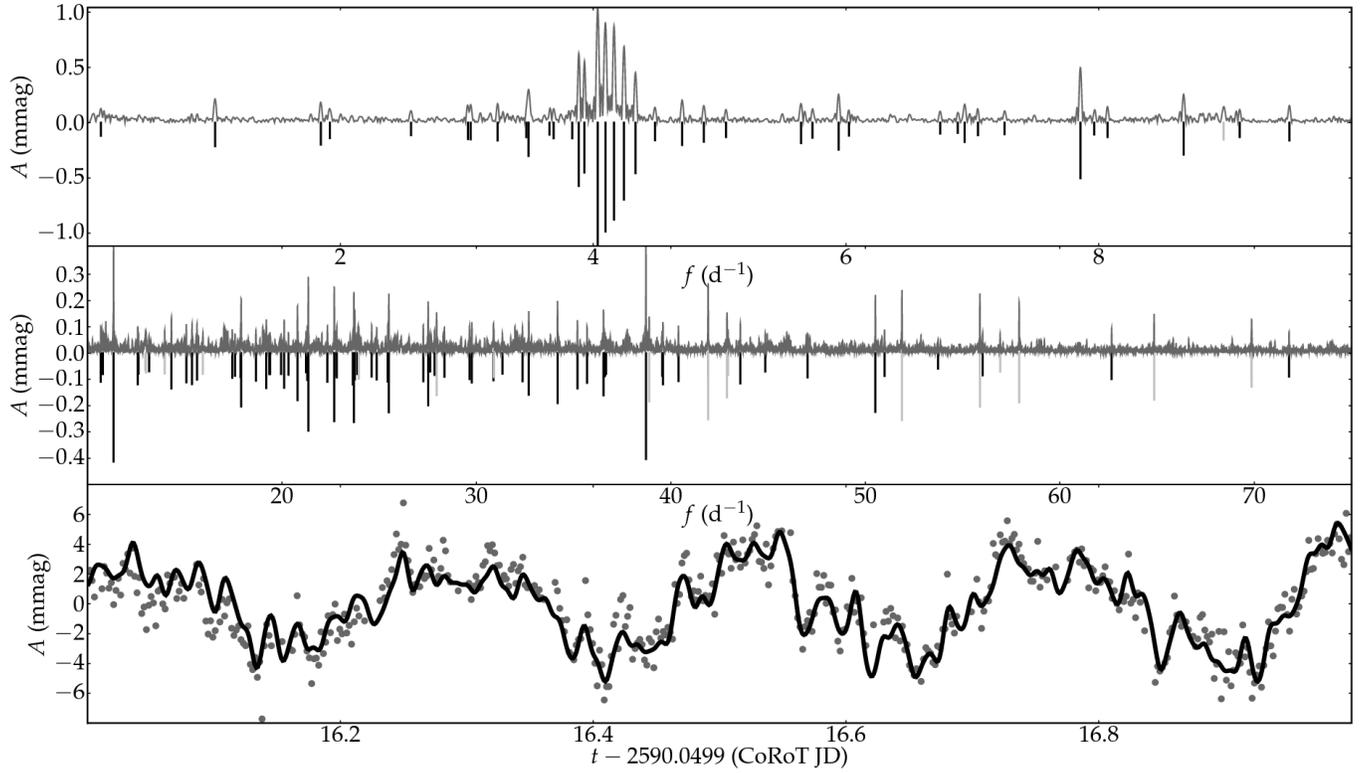}
\caption{$\beta$\,Cep candidate 102729531: (\emph{bottom panel}) Zoom on day
17 of IRa01: on top of the clear $\sim0.25$\,d period, many more significant
variations of the flux are noticable. (\emph{upper two panels}) raw frequency
spectrum (grey) with identified frequencies with SNR$>4$ (black lines) and
orbital frequencies (grey lines, all harmonics and day aliases of 1 to 5
days).}\label{fig:app:rp_BCEP_102729531}
\end{figure*}

\begin{figure*}[htb]
\centering\includegraphics[width=2\columnwidth]{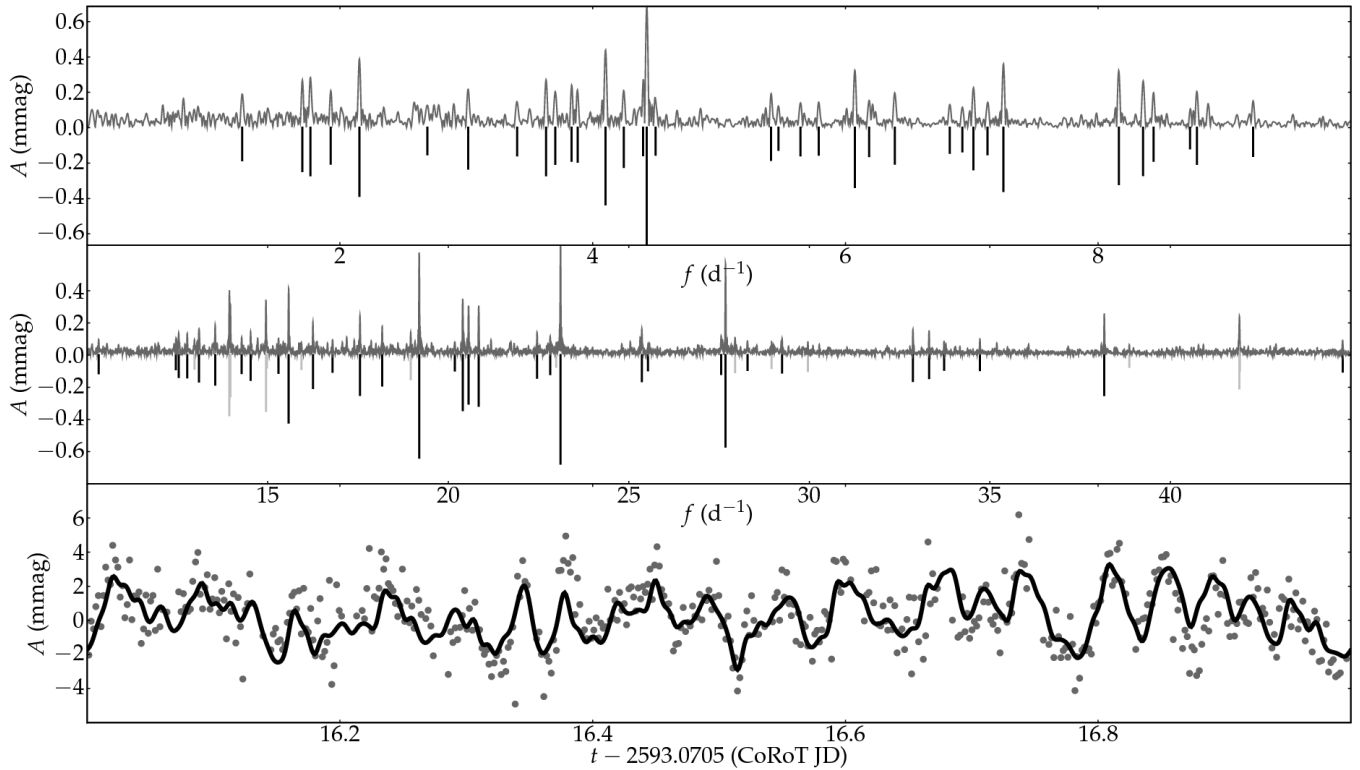}
\caption{Same as Fig.\,\ref{fig:app:rp_BCEP_102729531}, but for the $\beta$\,Cep
candidate 102790063.}\label{fig:app:rp_BCEP_102790063}
\end{figure*}

\clearpage
\begin{figure}
\includegraphics[width=\columnwidth]{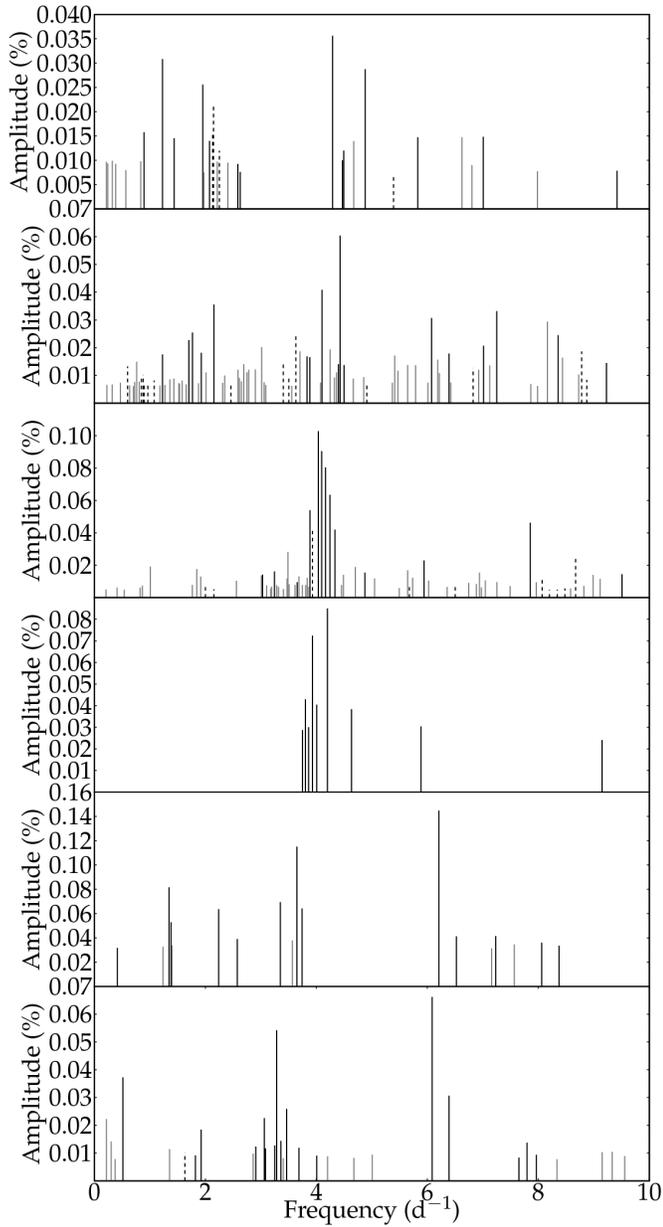}
\caption{Schematic oscillation spectra of one group of pulsators of which the fundamental parameters $\log T_{\rm eff}$ and $\log g$ place them in between the SPB and $\delta$\,Sct
instability strip (CoRoT 102861067, 102790063, 102729531, 102862454, 102790331, 102771057). These are all multiperiodic variables, showing complex
variability in a broad frequency range (independent modes are depicted with black solid lines, harmonics as interrupted black lines,
and candidate combination frequencies in grey).\label{fig:app:complex_freqbar}}
\end{figure}

\begin{figure}
\includegraphics[width=\columnwidth]{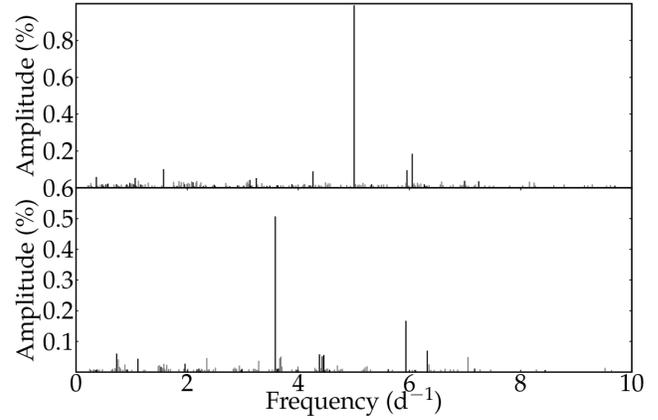}
\caption{Same as Fig.\,\ref{fig:app:complex_freqbar}, but for two of the stars with a rich frequency spectrum,
but having few large-amplitude pulsations (CoRoT 102850502, 102816758).\label{fig:app:highampl_freqbar}}
\end{figure}

\begin{figure*}[htb]
\centering\includegraphics[width=2\columnwidth]{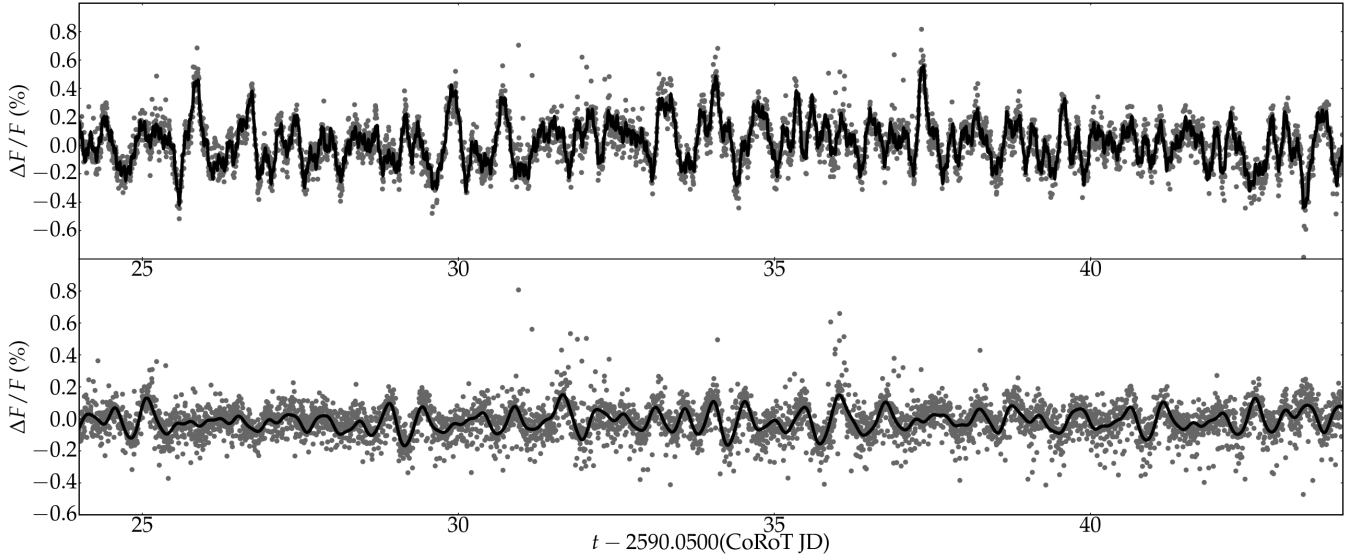}
\caption{CoRoT 102872265 (\emph{upper panel}) and 102832493 (\emph{lower panel}) show a complex frequency
spectrum with high peaks with frequencies above the typical SPB pulsation range. Moreover,
the dense spectrum makes it difficult to seperate long, stable pulsations from beating patterns.
The solid line is a fit, grey circles are the data.}\label{fig:app:erratic_motions_spb}
\end{figure*}

 \begin{figure}[p]
\centering\includegraphics[width=\columnwidth]{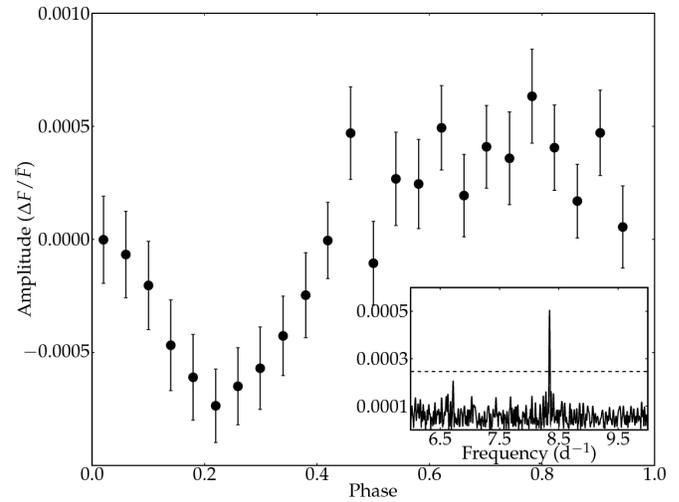}
 \caption{CoRoT 102808565: Binned phase diagram (1$\sigma$ errors on average),
 after removal of a nonlinear fit with 69 frequencies.  Inset: excerpt from the
 residual Scargle periodogram.}\label{fig:app:corot_102808565_highfreq}
 \end{figure}

 \begin{figure}[p]
\centering\includegraphics[width=\columnwidth]{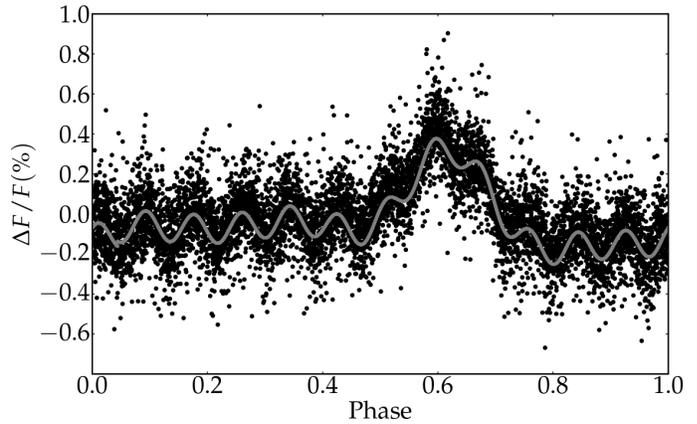}
 \caption{CoRoT 102762284: Phase diagram of the residual lightcurve after removal of the first two frequencies, which are both
well approximated by sinusoidal variations. The diagram was folded with frequency $f=0.27812$\,d$^{-1}$. A smaller period variation is noticable,
as well as an outburst-like peak in the second half of the phase.}\label{fig:app:outburst}
 \end{figure}

\begin{figure*}[p]
\centering\includegraphics[width=0.85\columnwidth]{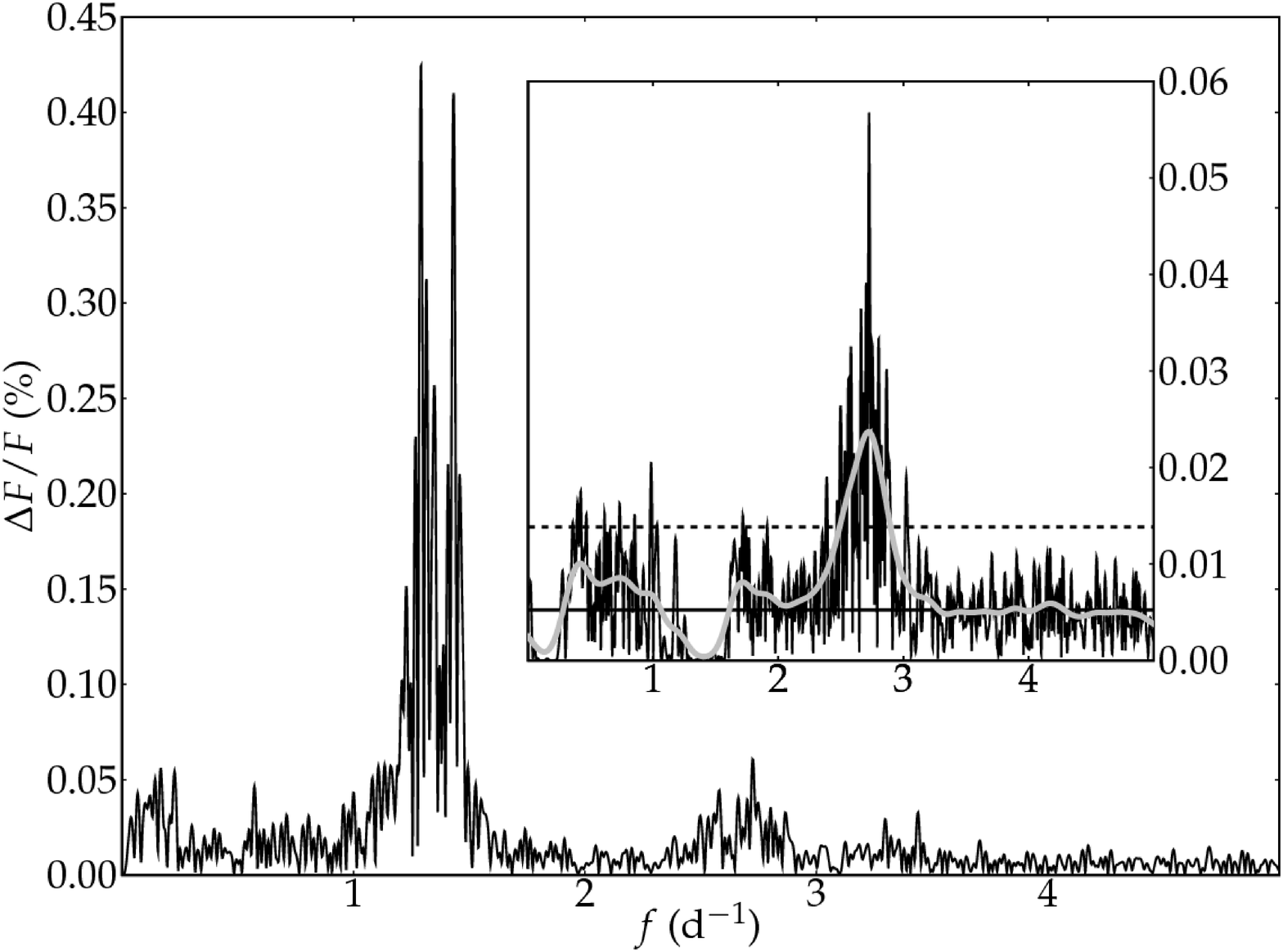}
\centering\includegraphics[width=0.85\columnwidth]{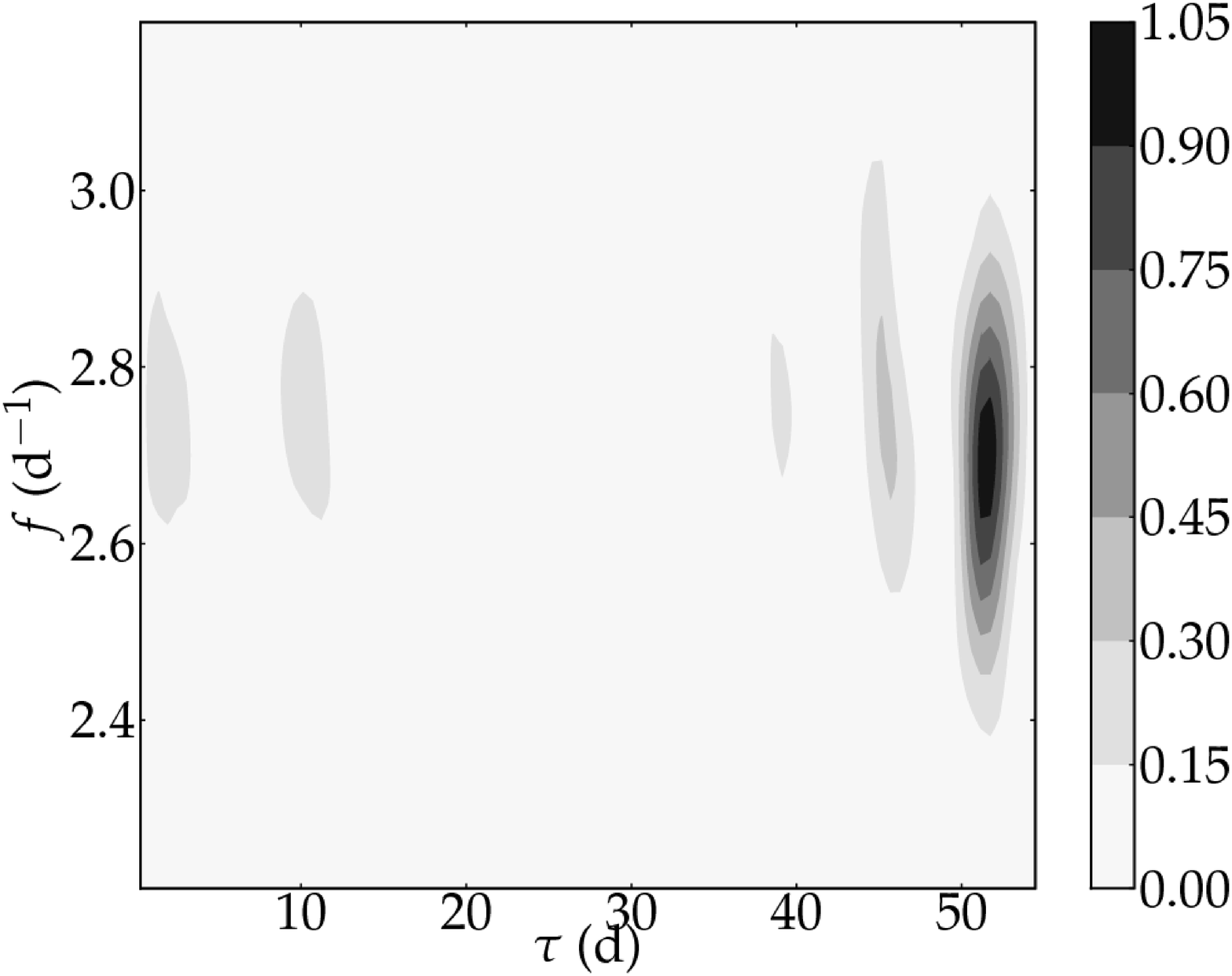}

\centering\includegraphics[width=0.85\columnwidth]{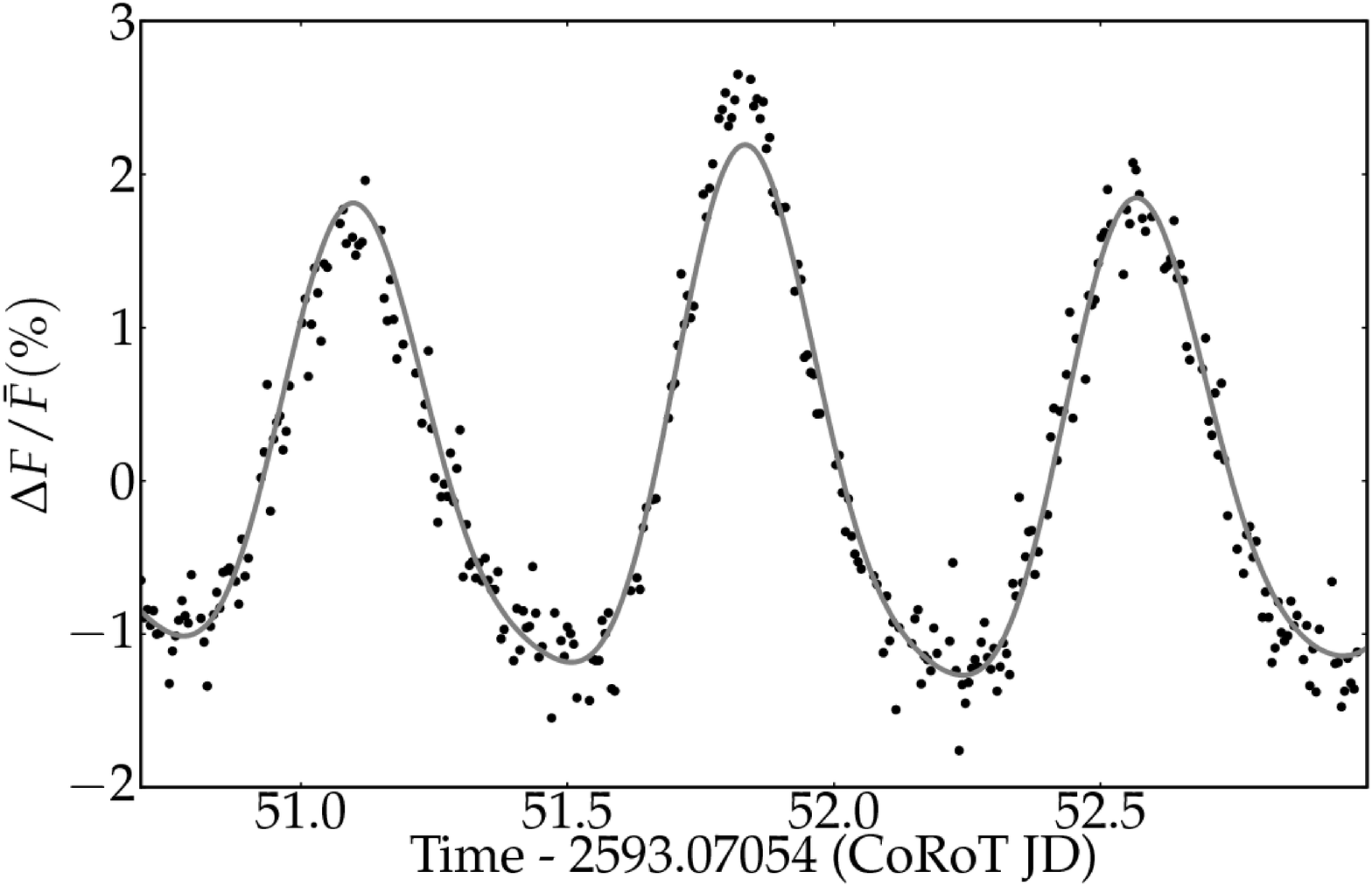}

\caption{CoRoT 102943966: Original periodograms of SPB candidates showing residual power excess at frequencies above $\sim$1.5 d$^{-1}$ (\emph{upper left}). The insets are amplitude spectra of the residuals, after prewhitening all frequency in lower regions. The grey solid line is a smoothed version of the spectrum, horizontal lines denote theoretical noise level and 99\% confidence interval. \emph{upper right:} Short time Fourier transformations of the residual light curves give the evolution of the (normalised) power in the periodogram over time since the start of the measurements ($\tau$). This time-frequency analysis shows that the excess power is due to modes with a finite lifetime in the order of days. (\emph{lower panel:}) excerpt from the light curve, showing the shape of the flux changes due to the pulsations.}\label{fig:app:power_excess1}
\end{figure*}

\begin{figure*}[p]
\centering\includegraphics[width=0.85\columnwidth]{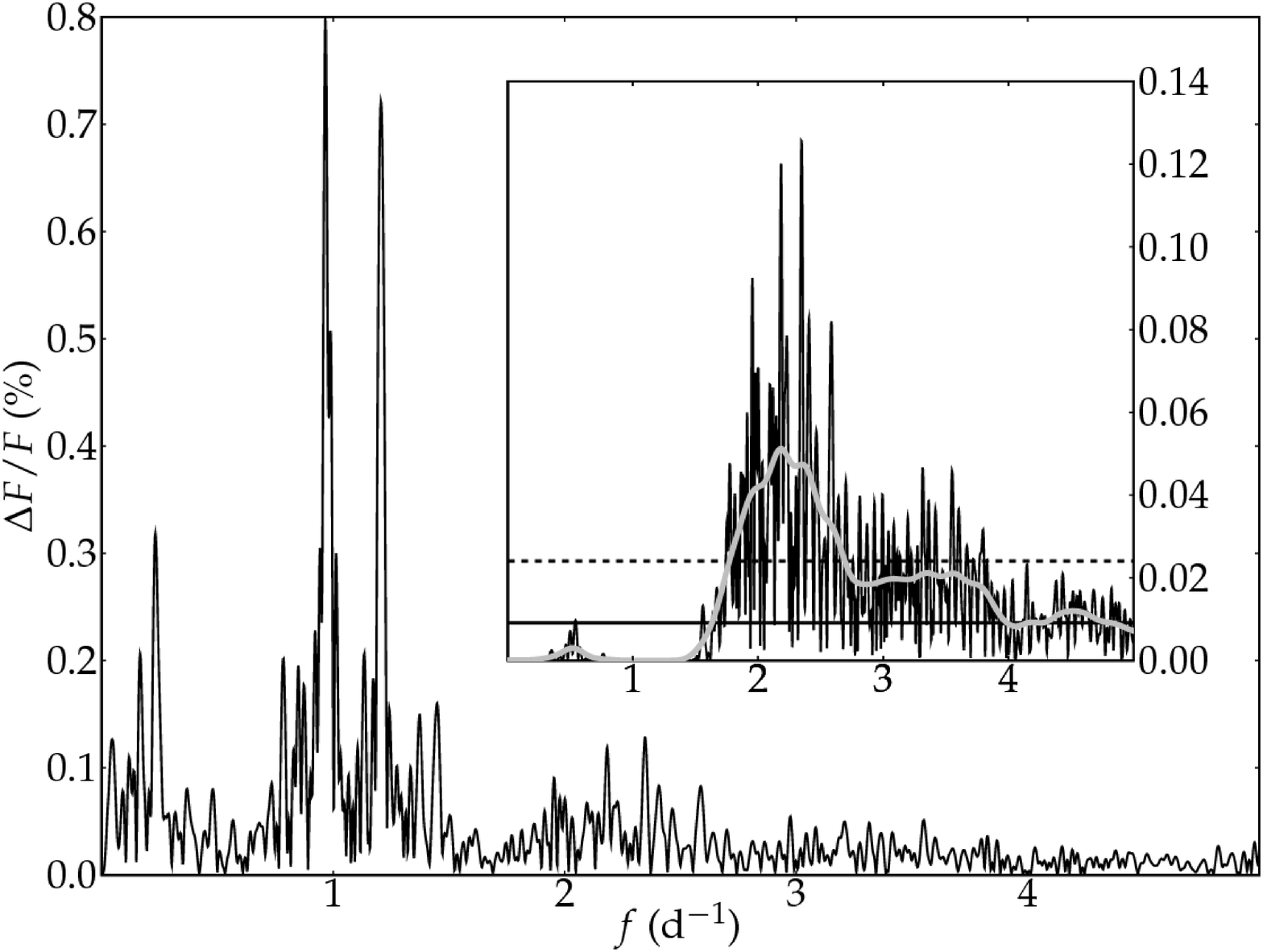}
\centering\includegraphics[width=0.85\columnwidth]{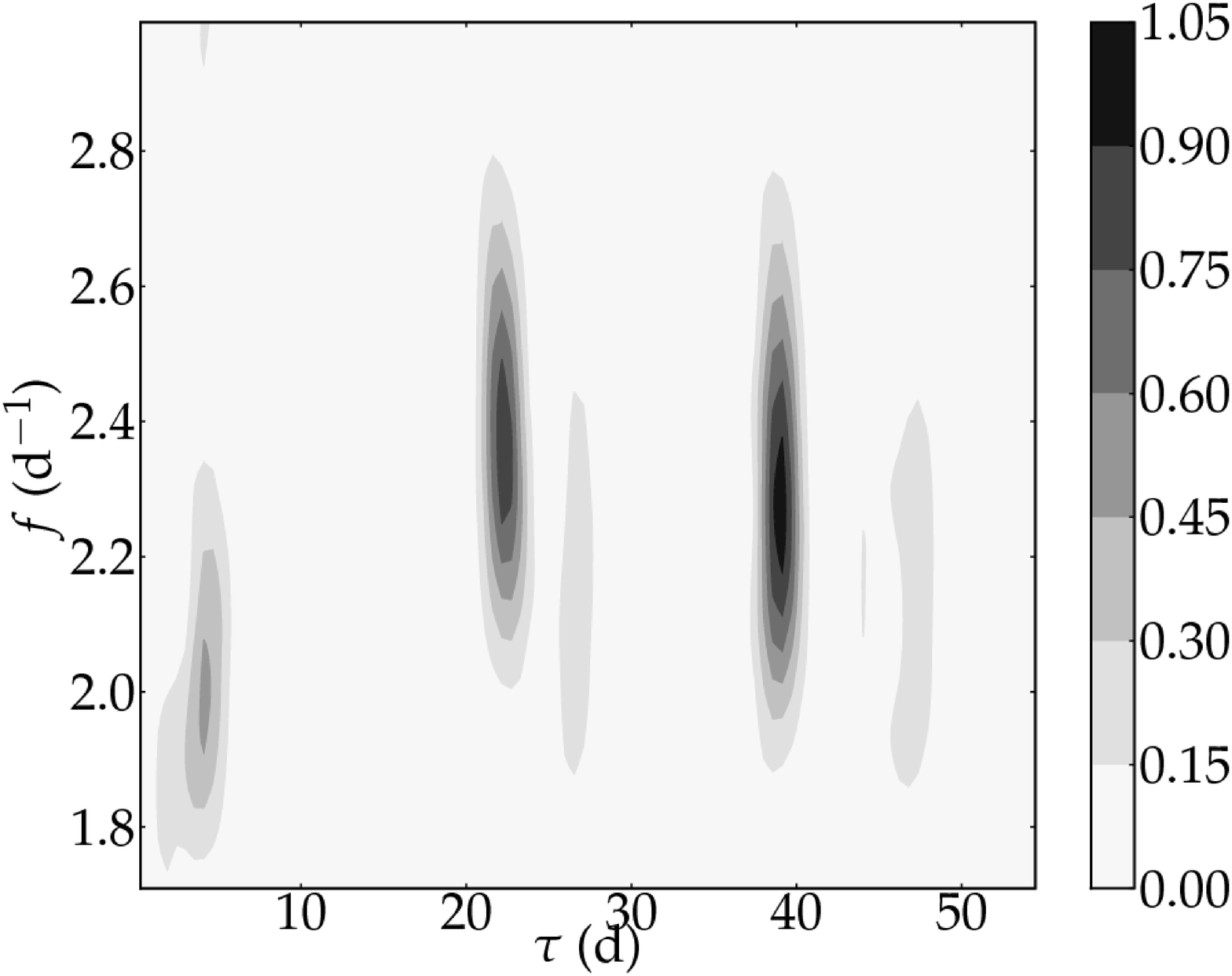}

\centering\includegraphics[width=0.85\columnwidth]{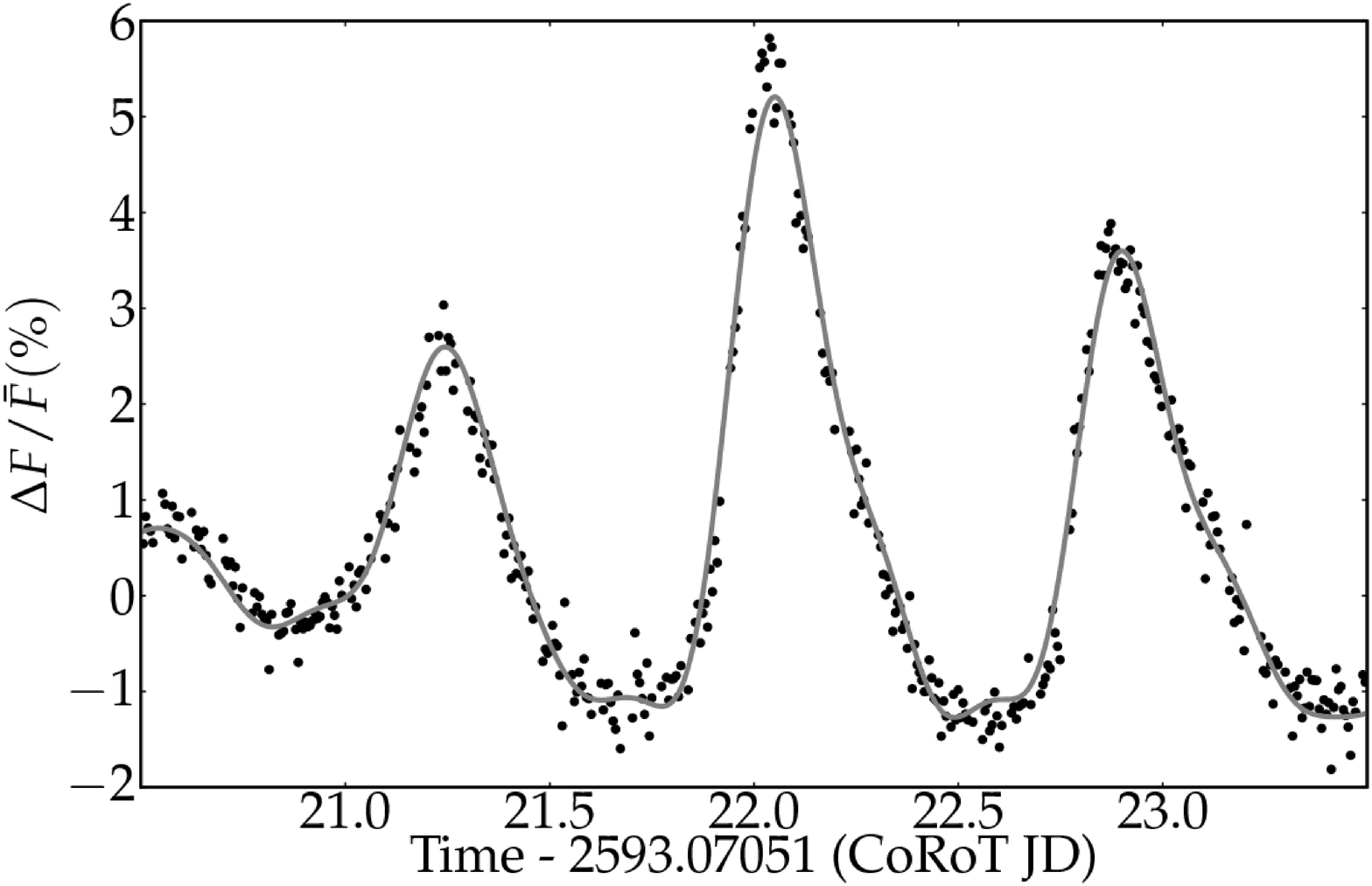}
\caption{CoRoT 102855391: same as Fig.\,\ref{fig:app:power_excess1}.}\label{fig:app:power_excess2}
\end{figure*}

\begin{figure*}[p]
\includegraphics[width=0.99\columnwidth]{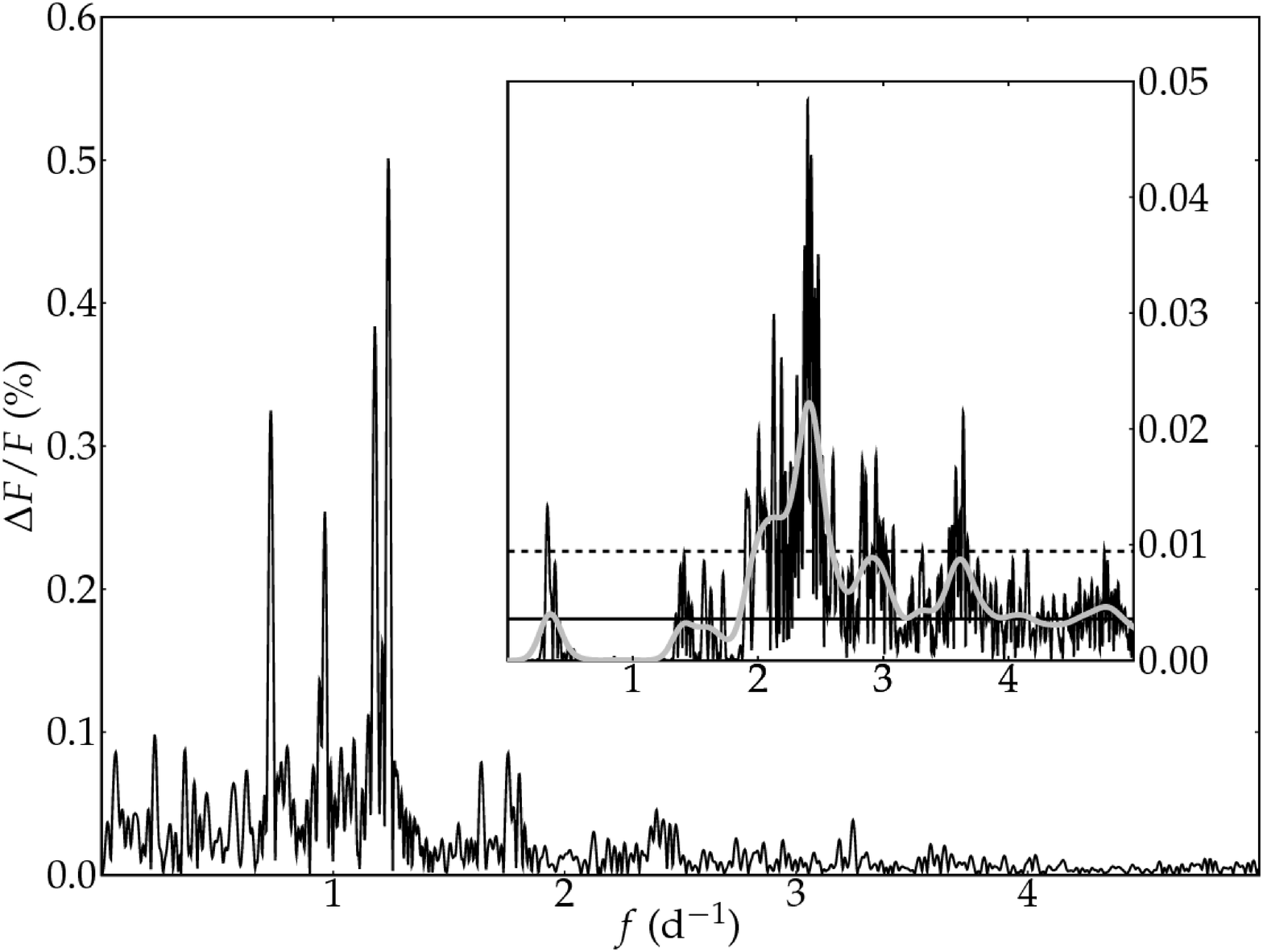}
\includegraphics[width=0.99\columnwidth]{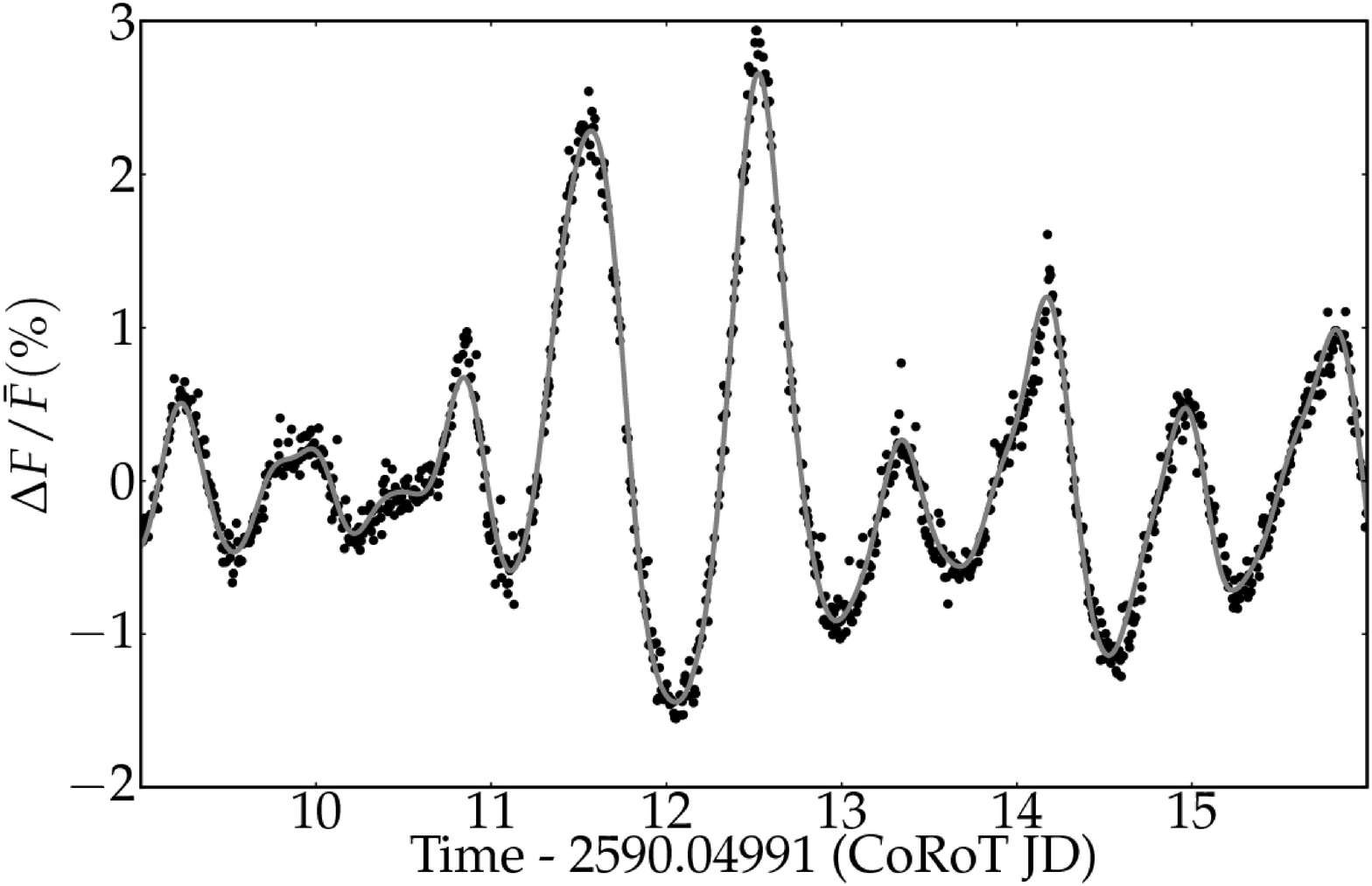}

\includegraphics[width=0.99\columnwidth]{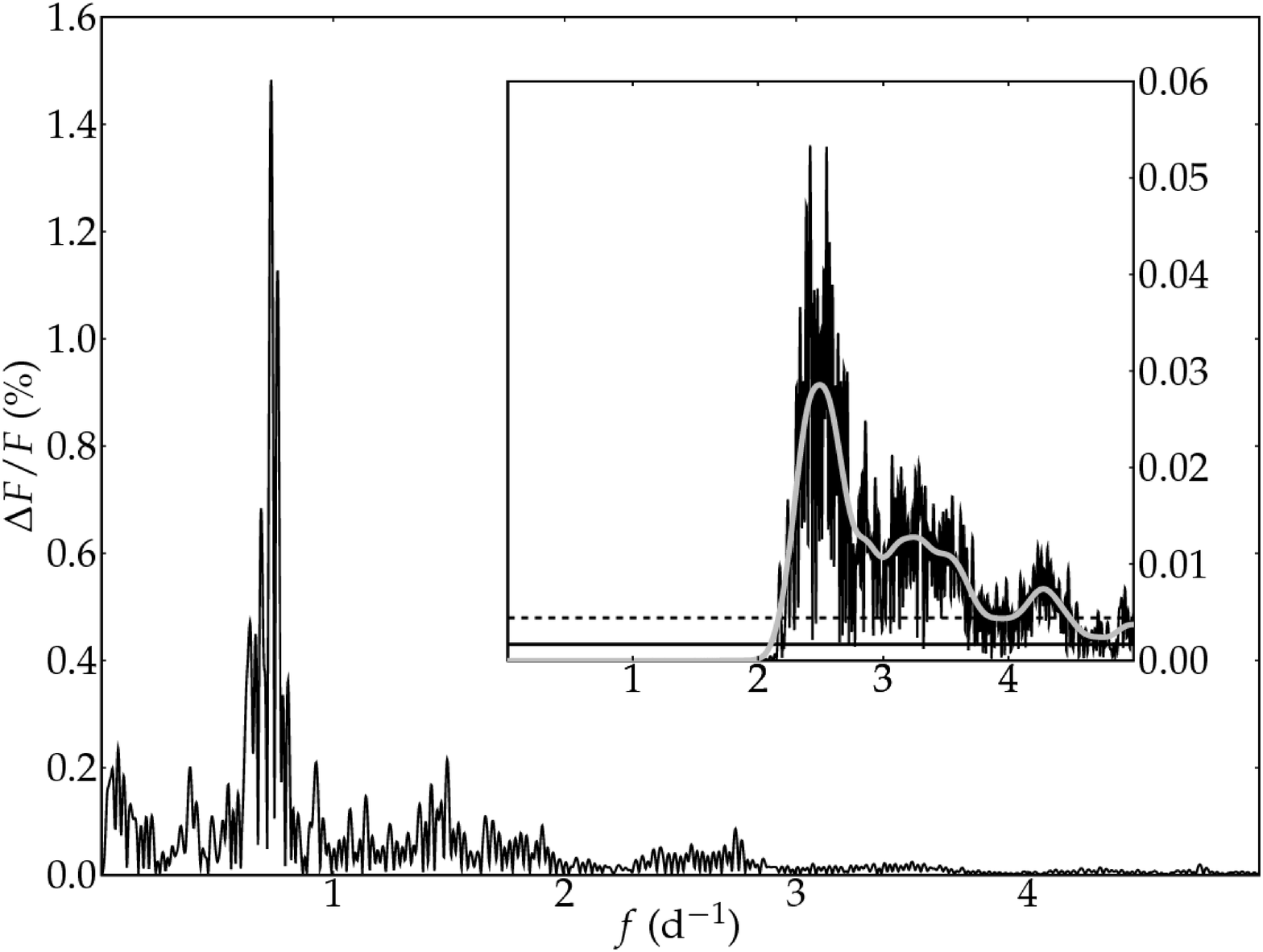}
\includegraphics[width=0.99\columnwidth]{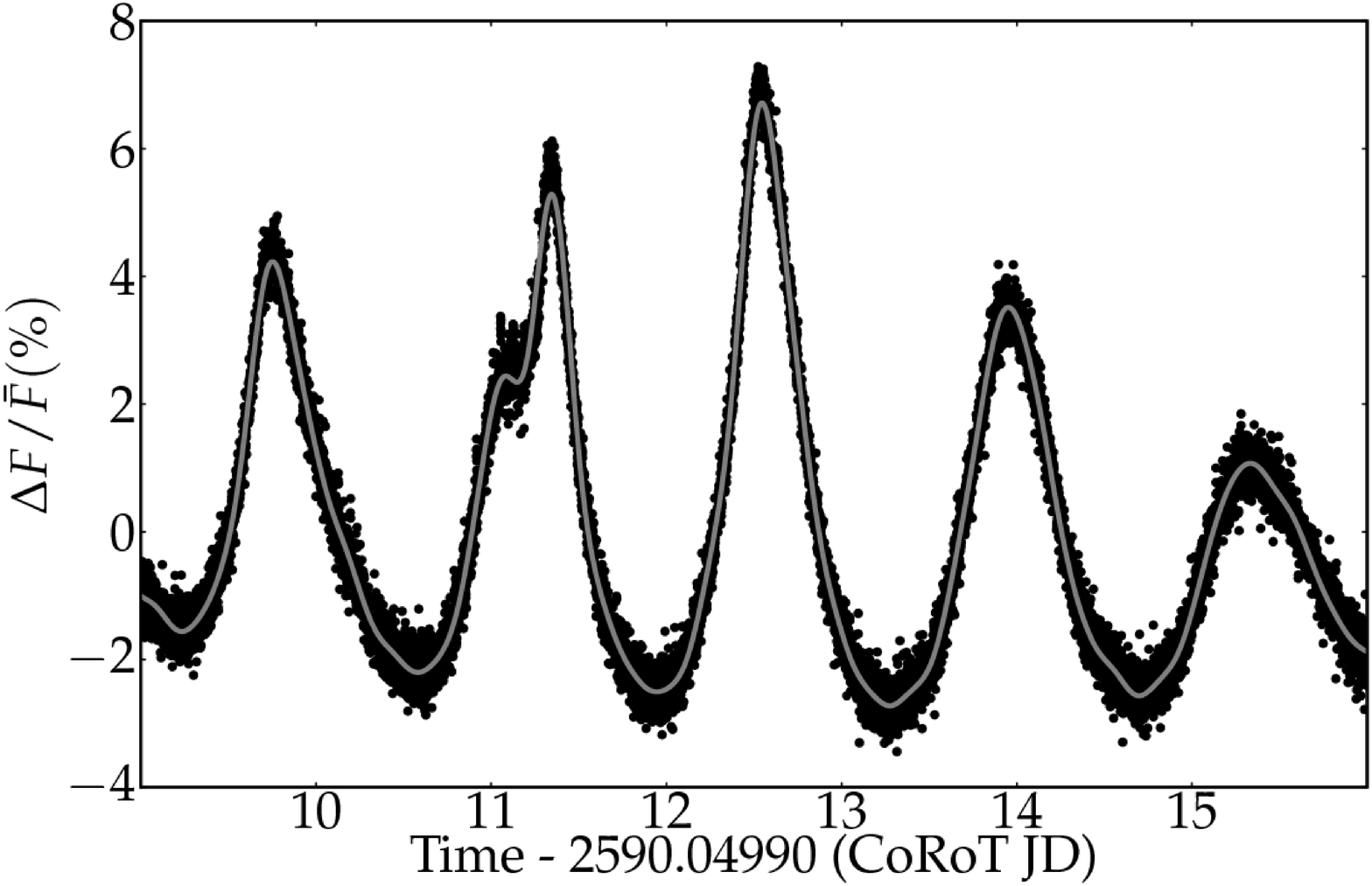}

\includegraphics[width=0.99\columnwidth]{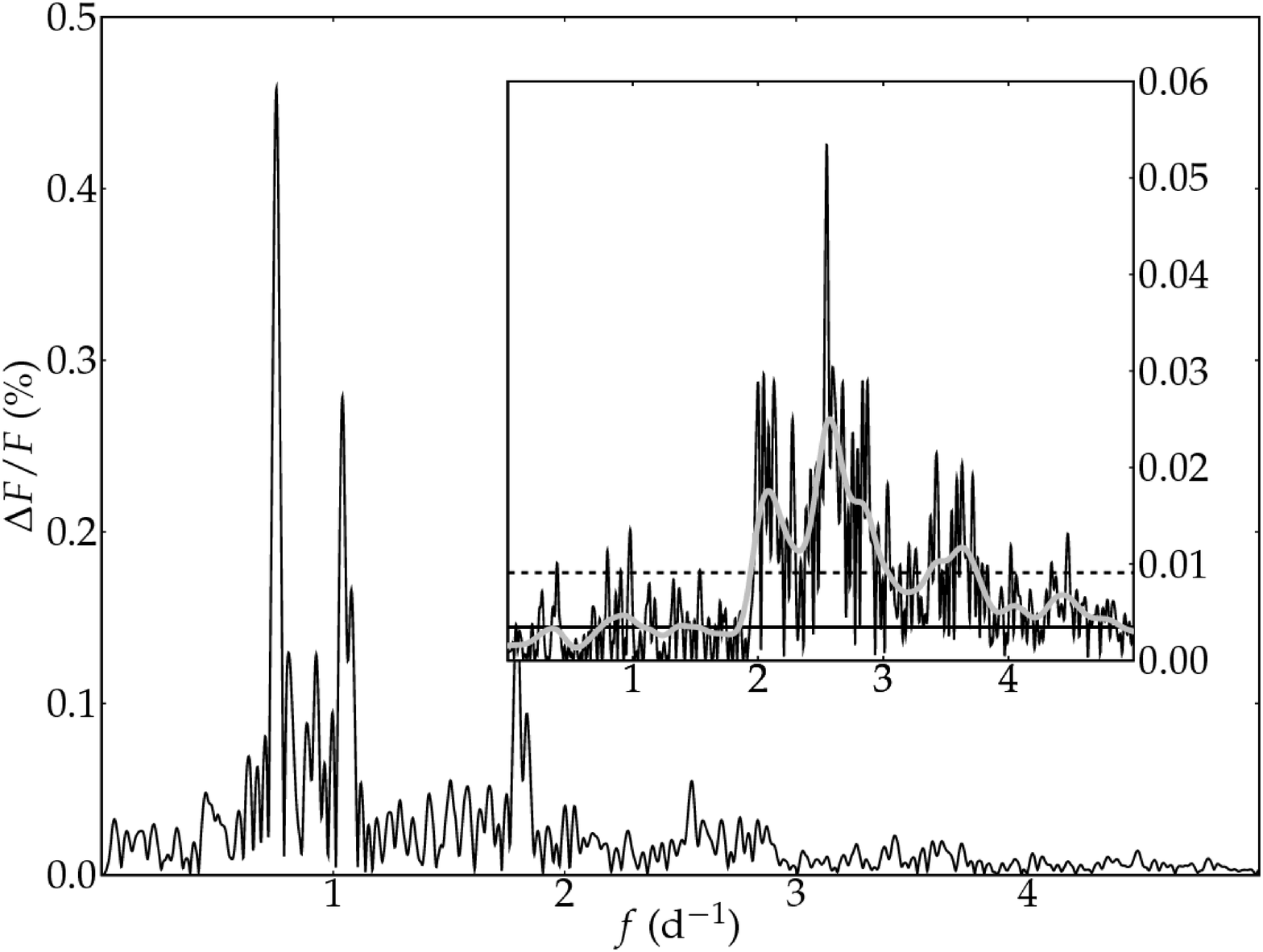}
\includegraphics[width=0.99\columnwidth]{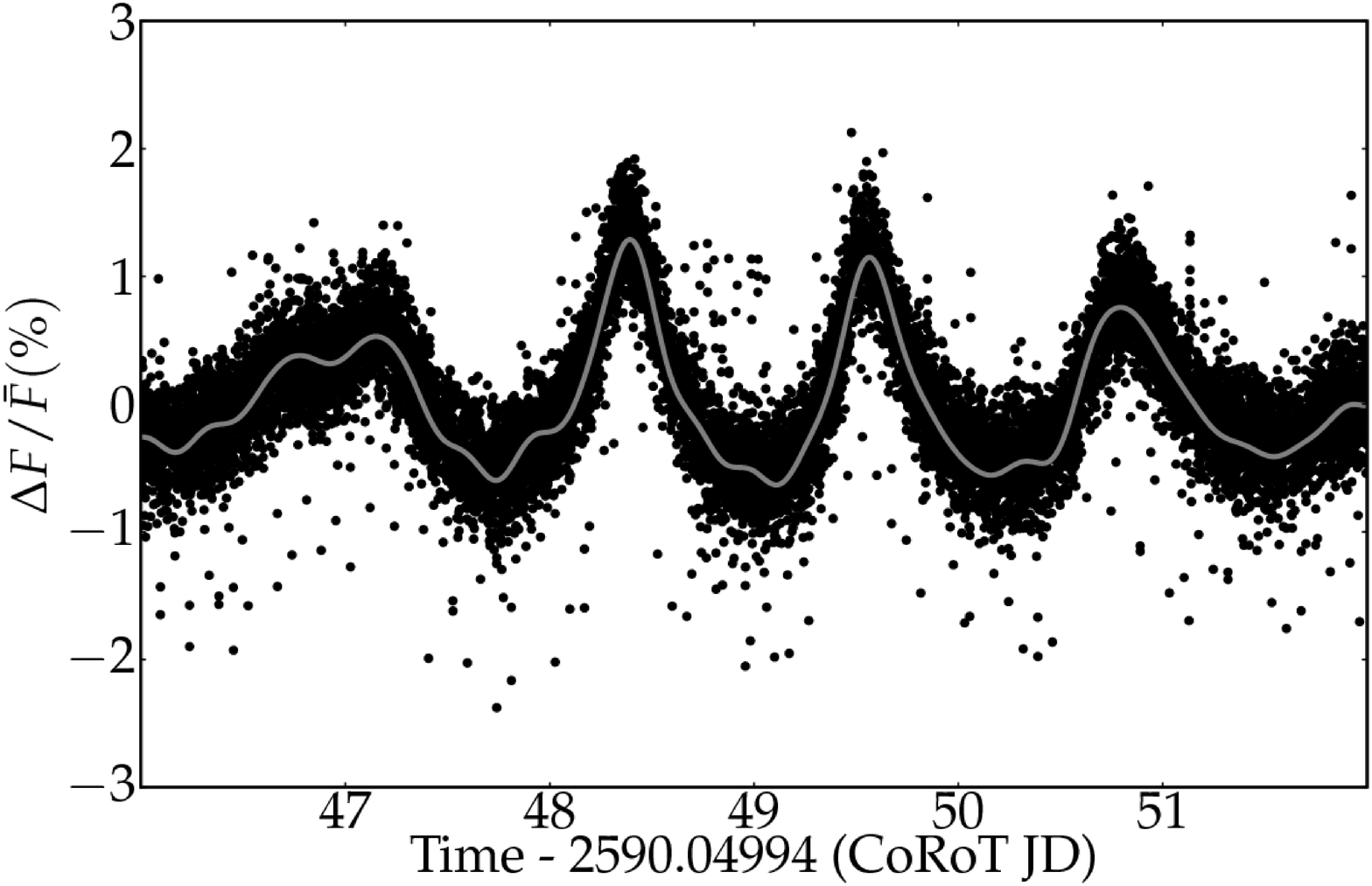}

\caption{(\emph{left column}) Original periodograms of SPB candidates showing residual power excess at frequencies above $\sim$1.5 d$^{-1}$. The insets are amplitude spectra of the residuals, after prewhitening all frequency in lower regions. The grey solid line is a smoothed version of the spectrum, horizontal lines denote theoretical noise level and 99\% confidence interval.  (\emph{right column}) excerpt from the lightcurve, showing the shape of the flux changes due to the pulsations. (\emph{from top to bottom}) CoRoT 102769848, 102754851 and 102848506.}\label{fig:app:power_excess4}
\end{figure*}

\begin{figure*}
\includegraphics[width=0.99\columnwidth]{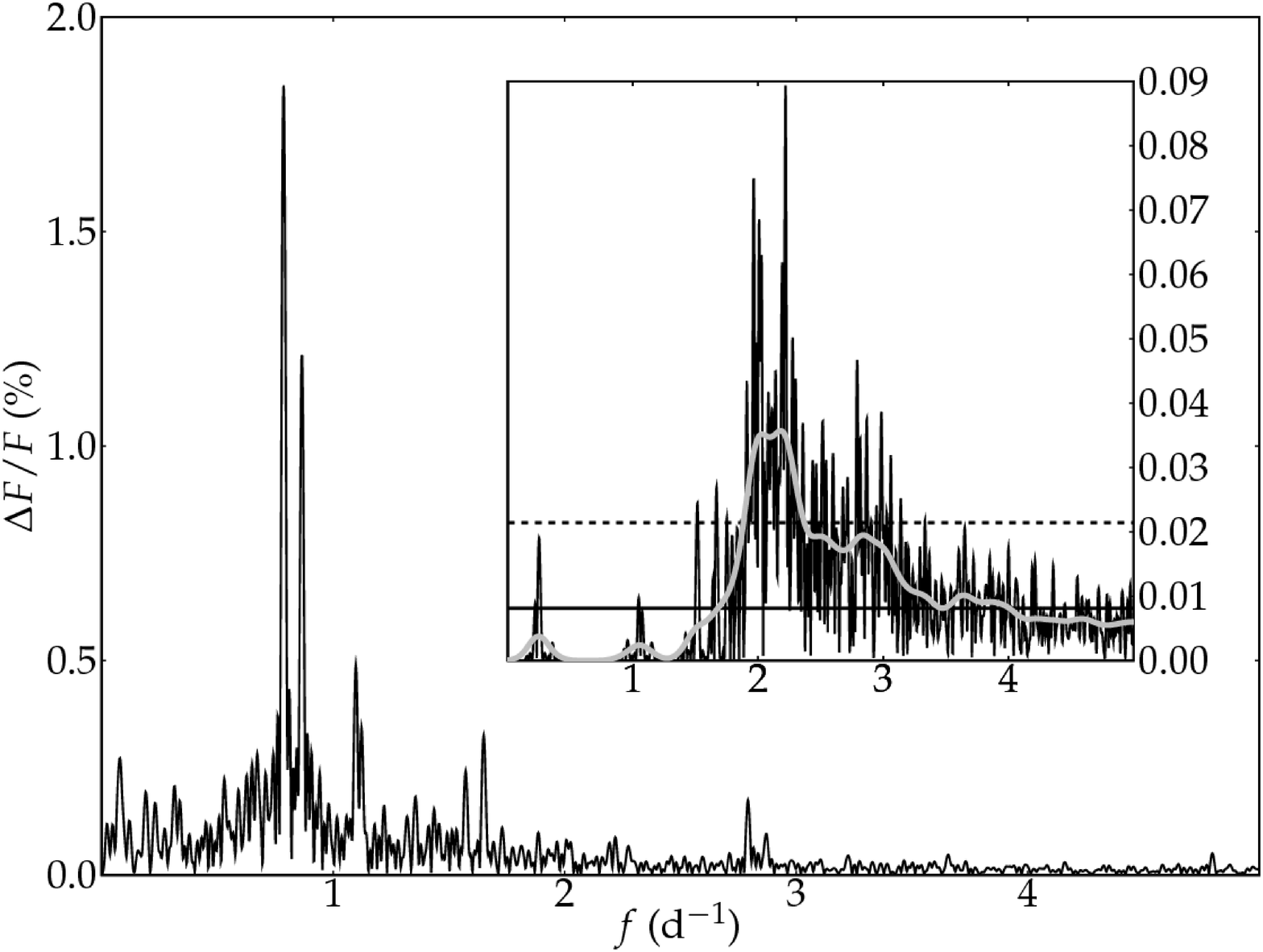}
\includegraphics[width=0.99\columnwidth]{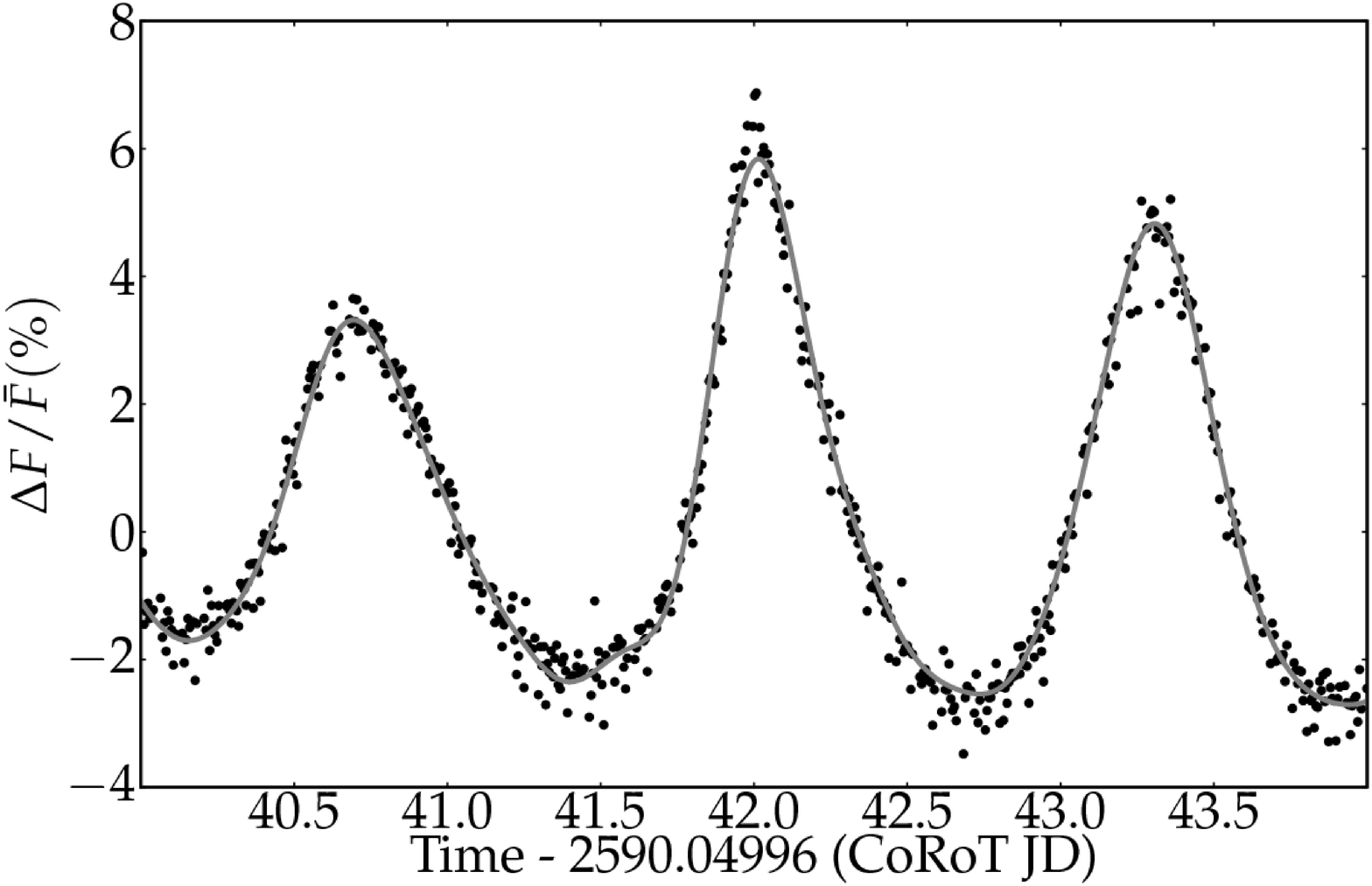}

\includegraphics[width=0.99\columnwidth]{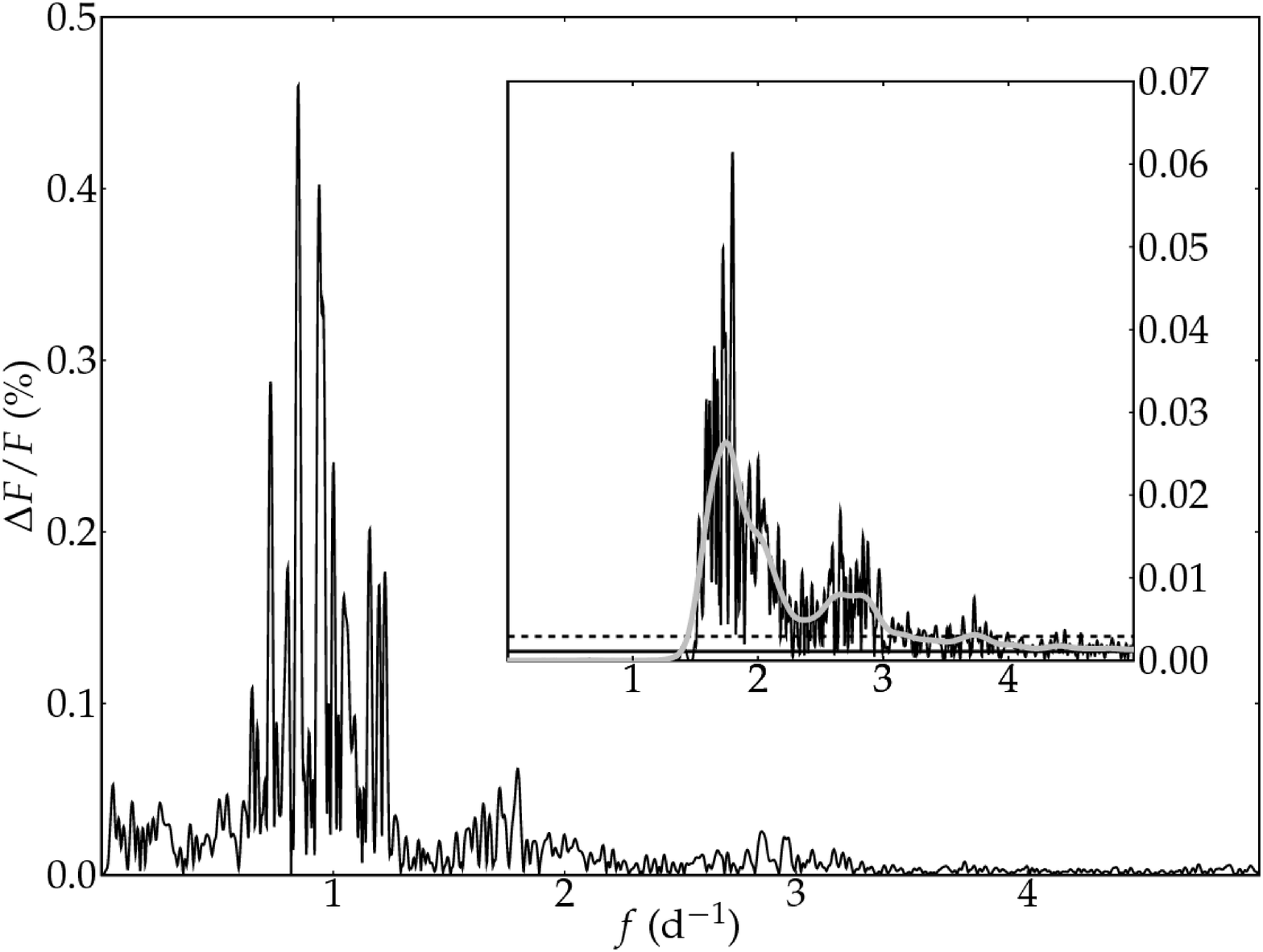}
\includegraphics[width=0.99\columnwidth]{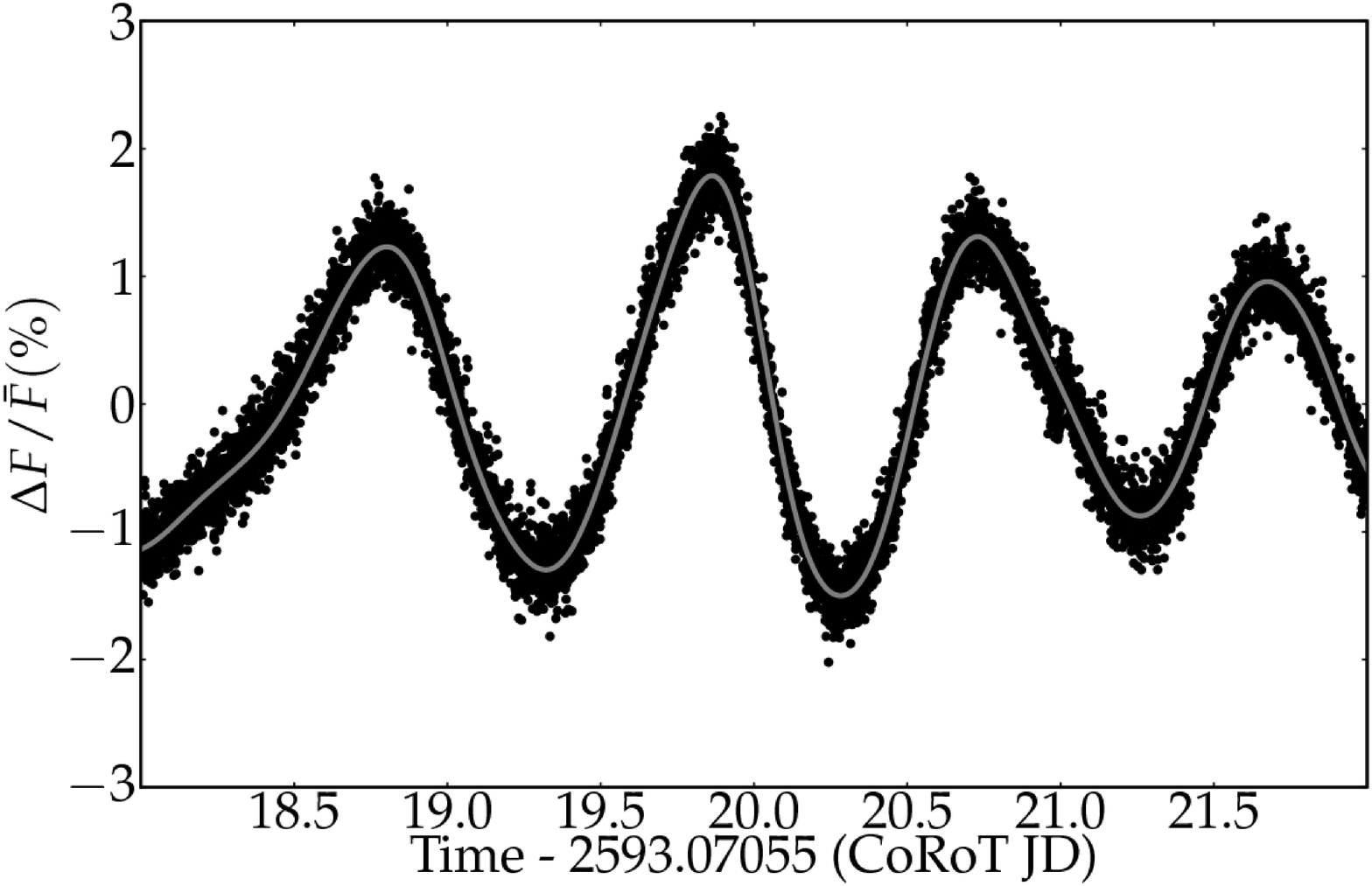}
\caption{Same as Fig.\,\ref{fig:app:power_excess4}, but for CoRoT 102844894 and 102956197.}\label{fig:app:power_excess5}
\end{figure*}

\begin{table}[p]
\caption{CoRoT 102804522 shows rich variability on varying timescales, but many of the frequencies can be connected through the same frequency spacing ($\epsilon=f_2-f_1-\Delta f$).}
\centering\begin{tabular}{lcccr}
\hline\hline Type & $f_1$\,(d$^{-1}$) & $\Delta f$\,(d$^{-1}$) & $f_2$\,(d$^{-1}$) & $\epsilon$\,(d$^{-1}$)\\\hline 
Single &   1.0891 & 0.5320 &  1.6213 & 0.0002\\
Single &   1.2350 & 0.5248 &  1.7595 &-0.0003\\
Single &   1.4445 & 0.5248 &  1.9697 & 0.0004\\
Single &   1.7595 & 0.5320 &  2.2914 &-0.0002\\
Single &   1.8758 & 0.5320 &  2.4077 &-0.0002\\
Single &   3.1081 & 0.5248 &  3.6325 &-0.0004\\
Single &   3.6325 & 0.5248 &  4.1572 &-0.0001\\
Single &   3.7455 & 0.5248 &  4.2703 &-0.0001\\
Single &   3.8578 & 0.5320 &  4.3899 & 0.0001\\
Single &   3.9212 & 0.5248 &  4.4458 &-0.0003\\
Single &   4.3259 & 0.5248 &  4.8512 & 0.0004\\
Single &   4.7765 & 0.5320 &  5.3088 & 0.0003\\
Single  &   9.0494 & 0.5248 &  9.5741 &-0.0001\\
Triplet &  14.2073 & 0.5320 & 14.7395 & 0.0001\\
        &  14.7395 & 0.5320 & 15.2719 & 0.0004\\\hline
\end{tabular}\label{tbl:spacing:102804522}
\end{table}

\end{appendix}
\end{document}